# Lab on a Chip

## CRITICAL REVIEW

# Advancing sustainable energy solutions with microfluidic porous media

Wenhai Lei,[a,b] Yuankai Yang,[c] Shuo Yang,[d] Ge Zhang,[e] Jenna Poonoosamy,[c] Anne Juel,[f] Yves Méheust,[d,g] Shervin Bagheri[a,b] and Moran Wang*[h]

The transition to a sustainable, low-carbon energy future requires transformative advancements in energy and environmental technologies. Carbon capture and sequestration, underground hydrogen storage, and nuclear waste geological disposal will be central aspects of a sustainable energy future, both for mitigating $CO_2$ emissions and providing green energy. A comprehensive understanding of multiphase flow through porous media, along with reactive transport and microbial activities, is essential for assessing the feasibility and managing the risks of these technologies. Microfluidic porous media platforms have emerged as powerful tools for the direct visualization of multiphase reactive flow in porous media and eventually optimizing these multiple physicochemical and biological processes. This review highlights critical scientific challenges associated with these sustainable energy solutions and summarizes the state-of-the-art microfluidic techniques for studying the interplay between multiphase flow, reactive transport, and biological effects in porous media. We provide a comprehensive overview of how these microfluidic approaches enhance the understanding of fundamental pore-scale dynamics and bridge the gap between pore-scale events and large-scale processes. This review is expected to promote both experimental and theoretical understanding of multiphase reactive flow in porous media, thereby informing material design, process optimization, and predictive modeling for scalable implementation. By fostering interdisciplinary collaboration across microfluidics, fluid mechanics, geophysics, materials science, and subsurface engineering, we hope to accelerate innovation and advance sustainable energy solutions.

## 1. Introduction

To achieve the 1.5-2°C global warming target outlined in the 2015 Paris Agreement, the Intergovernmental Panel on Climate Change (IPCC) emphasizes the urgent need for multifaceted solutions, including the widespread adoption of renewable energy, exploring effective carbon reduction and utilization approaches, and the scaling of carbon capture and storage technologies. At the current stage, a direct transition from fossil fuels to carbon-free energy sources is insufficient to meet wide energy demand. The International Energy Agency (IEA)'s Net Zero by 2050 report stresses that energy systems must undergo unprecedented changes, as the energy sector contributes roughly 75% of global greenhouse gas emissions[1]. In this case, large-scale carbon capture and sequestration (CCS), underground hydrogen storage (UHS), and nuclear energy are critical strategies under exploration to achieve global climate control goals[2]. By 2050, capturing and storing 3-10 gigatons of $CO_2$ annually - comparable to the current fluid-handling capacity of the petroleum industry - will be necessary to mitigate the most severe impacts of climate change[2]. UHS, still in development, offers a complementary approach to address seasonal fluctuations in renewable energy sources such as solar and wind[3]. Moreover, nuclear power is also regarded as a sustainable energy source with the potential to mitigate global warming. However, its adoption is contentious in many countries due to concerns over nuclear waste management[4]. Effective nuclear waste geological disposal (NWGD) is essential to ensure long-term environmental protection and safety for future generations. These needs for CCS, UHS, and NWGD underscore their significance in reducing emissions and stabilizing renewable energy supply. While the fundamental mechanisms of various sustainable energy systems—such as advanced materials, fuel cells, electrolysis for $CO_2$ reduction and hydrogen generation, and geothermal energy extraction—may share commonalities, these will not be discussed in detail here. To evaluate the feasibility and manage the risks of these sustainable energy solutions, it is essential to understand and predict multiphase reactive flow through these porous media. Yet, the mechanisms governing multiphase flow, reactive transport, and microbial activities in these porous environments remain poorly understood. This is because the

[a.] FLOW Centre, Department of Engineering Mechanics, KTH Royal Institute of Technology, Stockholm 100 44, Sweden.
[b.] Wallenberg Initiative Materials Science for Sustainability (WISE), Stockholm 100 44, Sweden.
[c.] Institute of Energy and Climate Research (IEK-6): Nuclear Waste Management and Reactor Safety, Forschungszentrum Jülich GmbH, Jülich 52425, Germany.
[d.] CNRS, Géosciences Rennes, University of Rennes, Rennes UMR 6118, France.
[e.] Department of Energy Science and Engineering, Stanford University, 367 Panama St., Stanford, CA 94305, USA.
[f.] Manchester Centre for Nonlinear Dynamics and Department of Physics and Astronomy, University of Manchester, Oxford Road, Manchester M13 9PL, UK.
[g.] Institut Universitaire de France (IUF), Rennes UMR 6118, France.
[h.] Department of Engineering Mechanics, Tsinghua University, Beijing 100084, China. E-mail: mrwang@tsinghua.edu.cn

underlying pore-scale events in porous media are more complex but are typically inaccessible to direct observation. Furthermore, the lack of a validated upscaling framework from the individual pore scale to the global porous system scale poses a significant barrier to linking interfacial phenomena to industrial applications.

Flow or transport phenomena in porous materials are often described using linear homogenized equations at the continuum/Darcy scale, which condense the significant deviations caused by multiscale heterogeneities - an inherent feature of transport in porous media - into a few effective parameters[5]. However, when these deviations interact with nonlinear behavior induced by coupled physicochemical and biological processes, the validity of such continuum scale descriptions becomes uncertain[6]. For example, multiphase reactive flows interact within the opaque porous medium, producing interfacial dynamics, mass transport, and dissolute or precipitate structures that span thousands to millions of pores[7]. Elucidating microscale multiphase flow dynamics and transport behavior and quantifying phase trapping or migration patterns in porous media is important for both physical understanding and industrial applications but remain challenging due to the difficulties of direct visualization and complex porous geometry. To visualize the multiphase reactive flow behavior and explore the underlying mechanisms in various porous media, microfluidic experiments and X-ray microtomography (micro-CT) are among the most popular methods[8]. Micro-CT imaging has been widely used for the pore-scale visualization of flow or transport in porous media. However, it has inherent limitations, such as challenges in achieving simultaneous imaging, high resolution, and rapid measurements, which remain under active investigation[9, 10]. Moreover, fabricating controllable and reproducible porous media remains a significant challenge, further limiting the application in controlled experiments with systematically varied parameters. Microfluidic experiments provide a convenient and precise approach for visualizing fluid flow, enabling detailed observations under controlled porous geometries and flow conditions. The advantages stem from the adoption of microfluidic chips (also known as "microfluidic porous media" or "micromodels"), which are transparent devices with characteristic length scales of less than a millimeter—designed for observing, managing, and manipulating fluid flow[11, 12]. The fabrication of microfluidic chips employs micro/nanofabrication techniques to transfer various functional designs onto specific wafers, enabling high-precision production and the creation of small-scale features, with the possibility to produce any two-dimensional geometry from numerical design. Microfluidic systems offer numerous advantages, including direct visualization, precise fluid manipulation, reproducible porous structures, and low analysis times. Their high spatiotemporal resolution and precise control of experimental conditions make them a powerful platform for investigating multiphase reactive flow dynamics in complex subsurface environments[13]. In the context of sustainable energy, microfluidic experiments not only enhance our understanding of the fundamental physicochemical and biological processes underlying key solutions outlined in the IEA roadmap[1]—such as carbon capture and sequestration and innovative methods for energy storage and extraction—but also serve as essential tools for exploring, developing, and testing new processes and materials critical to achieving these goals[14]. Insights gained from pore-scale observations can be upscaled to inform large-scale models and applications, driving progress in sustainable energy solutions. As the energy industry transitions away from crude trial-and-error approaches, microfluidic systems are poised to play a critical role by providing physics-based methodologies for studying multiphase reactive flow. By bridging the gap between microscale observations and real-world applications through physical upscaling techniques, microfluidic systems offer invaluable insights for the design and optimization of sustainable energy solutions.

In this review, we begin by delineating key scientific challenges associated with sustainable energy solutions, including carbon capture and sequestration (CCS), underground hydrogen storage (UHS), and nuclear waste geological disposal (NWGD). Subsequently, we summarize state-of-the-art microfluidic techniques to study multiphase reactive flow in porous media. To understand the fundamental behaviors underpinning these sustainable energy applications, we systematically review the various physicochemical and biological processes at the pore scale and explore how these mechanisms influence macroscopic outcomes. Finally, we present upscaling frameworks that connect fundamental research in confined multiphase reactive flow with practical applications in material science, subsurface engineering, and sustainable energy. Our goal is to inspire the microfluidic research community to explore the abundant opportunities emerging in the energy transition. This review aims to bridge pore-scale multiphase reactive flow research with large-scale applications, fostering interdisciplinary collaboration and underscoring the pivotal role of microfluidics in tackling energy transition challenges.

## 2. Key scientific problems for sustainable energy solutions

This review focuses on multiphase flow in key sustainable energy applications, including carbon capture and sequestration (CCS), underground hydrogen storage (UHS), and nuclear waste geological disposal (NWGD). Despite significant differences in fluid properties, storage environments, and operational approaches—such as the permanent containment goals for $CO_2$ storage and nuclear waste disposal, versus the temporary storage and demand-based extraction of hydrogen—these systems share common flow and transport mechanisms (Fig. 1). Enhanced understanding of multiphase flow in porous media, which encompasses complex physical, chemical, and biological processes, is crucial to the advancement of these sustainable energy technologies (Fig. 1). By consolidating this knowledge, we can accelerate the development of these sustainable energy technologies. This section delves into the fundamental scientific questions underpinning these solutions, which represent some of the most critical and widespread challenges in the field.

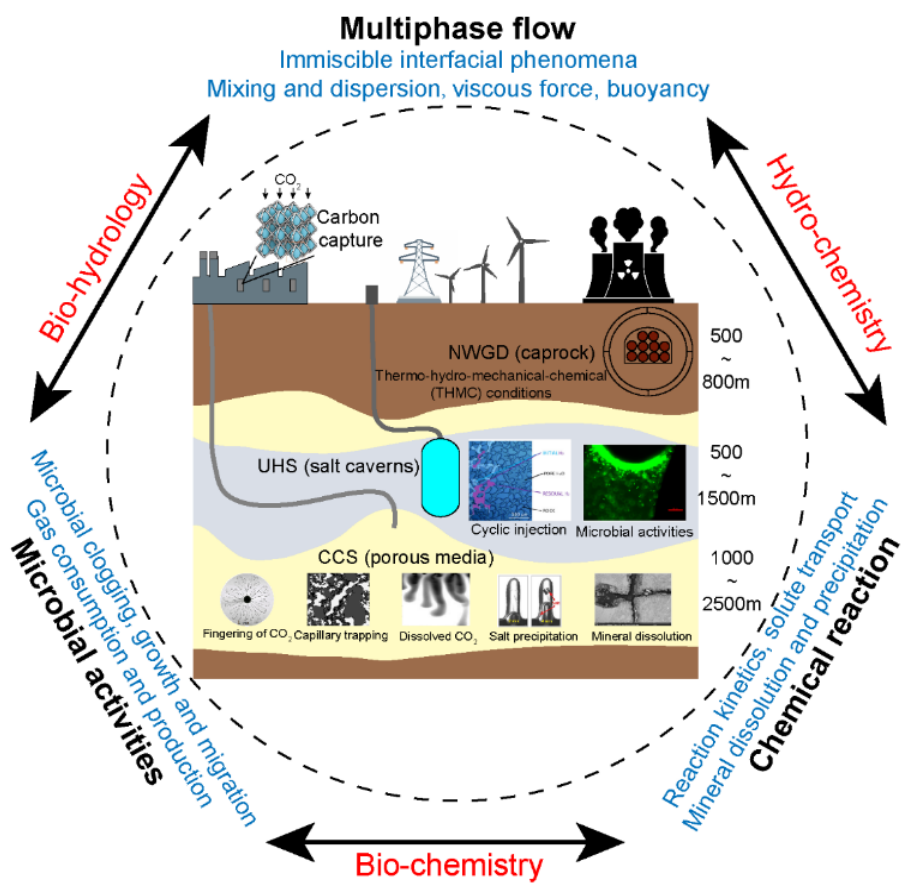

**Fig. 1.** Multiphase reactive flow in sustainable energy solutions (CCS, UHS, NWGD), coupling the multiphase flow, reactive transport, and microbial activities. The internal schematic diagram is a comparison between the subsurface storage of $CO_2$, $H_2$, and nuclear waste. The illustrations for CCS from left to right are viscous fingering[15], capillary trapping[16], $CO_2$ dissolution[17] (Reproduced with permission, Copyright 2006, Cambridge University Press), salt precipitation[18] (Reproduced with permission, Copyright 2013, Royal Society of Chemistry), and mineral dissolution[19] (Reproduced with permission from the author). The illustrations for UHS from left to right are cyclic injection[20] (Reproduced with permission under a CC-BY Creative Commons license, Copyright 2023, Elsevier), and microbial activities[21] (Reproduced with permission under a CC-BY Creative Commons license, Copyright 2023, Royal Society of Chemistry). The illustration for carbon capture is liquid-infused material[22].

## 2.1 Carbon capture and sequestration (CCS)

### I. CO2 capture

$CO_2$ can be captured from large sources, such as power plants, natural gas processing facilities, other industrial processes, and even the open atmosphere. Considering that anthropogenic $CO_2$ emissions have the largest impact on climate change, with more than 40% of global $CO_2$ emissions originating from coal-fired power plants[2, 23], economically feasible $CO_2$ removal from power plant flue gases presents one of the greatest challenges of our time. There are three techniques to remove or 'scrub' $CO_2$ where fossil fuels are burnt at power plants: post-combustion, pre-combustion, and oxyfuel combustion. Chemical absorption, particularly using aqueous amine solutions, has been the primary method for $CO_2$ removal from power plant emissions for decades[24]. Although still the only commercially viable technique, liquid amines work as scalable and high-density active agents for $CO_2$ capture through chemical interactions[25]. However, current industrial practices use large absorption columns with a limited gas-liquid surface area-to-volume ratio (A/V) of approximately 500 m$^{-1}$, which restricts the efficiency of gas-liquid interactions[26]. To reduce viscosity effects, amines are typically diluted with large amounts of water. However, the high specific heat capacity and heat of vaporization of water significantly increase energy demand during solvent regeneration. Furthermore, water dilution limits $CO_2$ capture at flue gas temperatures above 100°C, requiring flue gas to be cooled below 60°C before absorption and the spent amine-water solution to be reheated above 110°C for $CO_2$ release. This process accounts for approximately 80% of the energy consumption during solvent regeneration, highlighting the need for more energy-efficient and cost-effective alternatives[27].

Recent innovations in $CO_2$ capture methods, including advanced membranes[28], amine-functionalized mesoporous adsorbents[29], and amine-appended metal-organic frameworks (MOFs)[30], have emerged but are hampered by higher costs, lower stability, and energy-intensive processes. Scaling these methods for industrial applications remains challenging. Liquid-based porous media technology, inspired by liquid-infused surfaces (LIS) or liquid-infused materials (LIM)[31], is a promising development that requires the transfer of single-phase porous systems to multiphase porous systems. These liquid-based porous media consist of a chemically functionalized microtextured solid substrate capable of trapping and stabilizing a liquid layer (LIS) or a liquid droplet (LIM) with millimetric to nanometric thick[22, 32, 33]. The infused liquid is strongly bound or trapped to the solid surface via capillary forces, and by controlling and structuring the underlying substrate, liquid layers can be shaped and structured, creating new technological opportunities. These systems are particularly advantageous for multiphase reactive flow processes, which can leverage the liquid infused into the porous structure as a reactive medium. These liquid-based porous systems present a promising pathway for energy-efficient and scalable $CO_2$ capture solutions.

**II. Geological carbon sequestration (GCS)**

Geological formations are highly effective and potentially practical for storing large volumes of $CO_2$ on a gigaton (Gt) scale[34, 35]. The long-term security of geological carbon sequestration (GCS) relies on effective immobilization and trapping mechanisms of $CO_2$ within the subsurface. In the overall $CO_2$ mitigation process, injection strategies significantly affect flow dynamics near the wellbore. As $CO_2$ migrates further, its movement is governed by rock and fluid characteristics, buoyancy forces, reservoir heterogeneity, and the geometry of stratigraphic traps. The propagation of $CO_2$ is primarily driven by pressure gradients originating from the injection or by its buoyancy arising from the density difference between $CO_2$ and brine, which in conjunction with reservoir and fluid properties, present both further challenges and opportunities for effective storage. The immobilization and trapping of $CO_2$ plumes are categorized into residual (capillary) trapping, solubility trapping, and mineral trapping, all of which are essential for ensuring long-term storage security[6, 34].

**Capillary trapping.** Capillary or residual trapping is considered one of the most reliable mechanisms for stabilizing $CO_2$ post-injection. As $CO_2$ plumes migrate, structural traps, such as geological domes, can contain buoyant $CO_2$ plumes[36]. Subsequent immobilization of the buoyant $CO_2$ plume can be driven by capillary or residual trapping, which becomes increasingly important under low capillary number regime and heterogeneous porous media[37]. In this process, capillarity is the dominant force governing the behavior of $CO_2$ ganglia in porous media, typically outweighing viscous, gravitational, and inertial forces[38]. Capillary trapping occurs within the confined pore spaces of the host rock[39] and is closely controlled by capillary pressure. The surface tension between $CO_2$ and brine, combined with the wettability of the rock in the process of $CO_2$ and brine, the geometric constraints, and topological evolution of the porous structure, results in complex ganglion dynamics within the reservoir and enables residual trapping to immobilize $CO_2$ effectively[40-42].

**Convective dissolution.** Dissolution occurs when $CO_2$ comes into contact with resident brine or water and can be significantly enhanced by mixing due to dispersion and flow through heterogeneous rock formations[6, 43]. This process accelerates dissolution rates in heterogeneous formations, and contributes, along with residual trapping, to stabilizing the $CO_2$ plume, thereby reducing leakage risks. The density of brine increases with rising $CO_2$ concentrations, reaching a maximum at approximately 3 wt.% under typical reservoir temperatures and pressures[44]. Consequently, the dissolution of $CO_2$ into brine creates density differences driven by $CO_2$ concentration gradients, which in turn can induce convective currents in the $CO_2$-saturated brine[43, 45, 46]. Similarly to thermally driven convection, dissolution-driven convection can dramatically enhance the rate of $CO_2$ dissolution, playing a crucial role in stabilizing sequestered $CO_2$[47]. In brine-filled carbonate formations, $CO_2$ dissolution often leads to acid production, which reacts with carbonate minerals to form wormholes and large channels, potentially compromising rock integrity[48]. Additionally, a small fraction of water (up to 5 wt.%) may dissolve into the supercritical $CO_2$ phase, potentially causing localized drying in areas with high $CO_2$ fluxes[49, 50]. This phenomenon has been observed to result in halite precipitation, which may reduce permeability[51] or decrease the mechanical strength of rock[52].

**Mineralization.** On much longer time scales, $CO_2$ may precipitate as carbonates, likely the ultimate form of stable trapping, $CO_2$ mineralization offers a long-term trapping solution by converting $CO_2$ into stable minerals[53]. This mechanism effectively reduces leakage risks through wells or faults. However, in sedimentary systems, this process is often limited by the scarcity of reactive minerals and the extremely large reaction time scale, which can amount to thousands of years[54]. Their efficiency depends on factors such as rock properties, leakage pathways, and the geometry of the storage complex. Reservoir heterogeneities that redirect flow away from potential leakage paths further enhance trapping. Comprehensive site assessments now evaluate not only the target reservoir and caprock but the entire storage complex to ensure safe and effective $CO_2$ storage.

**2.2 Underground hydrogen storage (UHS)**

Despite decades of extensive research on $CO_2$ storage, the unique physical, chemical, and molecular properties of hydrogen ($H_2$) introduce a distinct set of challenges[55]. Underground hydrogen storage (UHS) is still in its early stages, with limited literature, few dedicated research efforts, and minimal large-scale testing. The lack of comprehensive laboratory data underscores the need for focused studies to understand the coupled processes governing $H_2$ storage in porous media[56-59]. UHS typically involves cyclic gas-liquid injection, comprising gas saturation (drainage) and desaturation (imbibition). Saturation hysteresis during this cycle affects saturation levels, directly impacting gas storage efficiency and extraction. $H_2$ also acts as an electron donor for anaerobic microorganisms, enabling them to reduce electron acceptors like $CO_2$ (for methanogens) or $SO_4^{2-}$ (for sulfate reducers), providing energy for their processes. At the pore scale, microbial activity can cause bio-clogging, altering rock wettability, hydraulic conductivity, and diffusivity, which affects storage performance. Unlocking the potential of UHS requires a better understanding of microbial conversion and its effects on storage performance. However, key questions about $H_2$ containment remain unresolved, such as how $H_2$ interacts with geologic seals, host rocks, brine, and fluids, leading to uncertainties in flow behavior and containment integrity. Addressing these gaps could enable the reuse of depleted hydrocarbon reservoirs and saline aquifers for $H_2$ storage. However, societal acceptance of UHS will depend on factors like sustainability, safety, transparency, and social equity[60]. The containment of hydrogen in subsurface reservoirs also depends on the integrity of caprocks and trapping mechanisms. While caprocks are proven barriers for $CO_2$ and methane ($CH_4$)[50, 61], their ability to prevent $H_2$ escape is uncertain due to hydrogen's smaller molecular size and higher diffusivity. Although the issue of leakage and caprock compatibility with hydrogen storage is important, it will not be covered here, as this review focuses on the challenges of multiphase reactive flow in porous media for UHS.

**Hysteresis effects.** The cyclic injection of hydrogen in porous media exhibits significant hysteresis, which refers to the differing relative permeability during gas injection and withdrawal processes. Since hydrogen is significantly less dense and less viscous than the formation fluid, it naturally spreads along the upper boundary of a layer upon injection, forming a narrow gravity-driven finger. However, during extraction, some water is drawn back to the well, reducing the recoverable hydrogen mass[62]. Over successive cycles, hydrogen accumulates within the formation until the amount extracted eventually equals the amount injected[63]. This phenomenon arises because the relative permeability of hydrogen and water during cyclic hydrogen injections is not constant and depends on the saturation history of the system, which complicates the hydrogen-water multiphase flow behavior[37]. These dynamic interactions are influenced by many factors such as saturation levels, pressure, and temperature. For instance, hydrogen storage capacity in pore spaces can reach up to 60% under cyclic injections, but the distribution of trapped hydrogen varies between cycles[20]. Therefore, the hysteresis effect poses a significant challenge for UHS because it affects storage efficiency and recovery rates. Addressing hysteresis requires both advanced experimental and modeling studies to minimize hydrogen losses and optimize storage performance[64].

**Microbial activities.** Microbial activity is a critical factor influencing $H_2$ loss in the subsurface[65]. A detailed understanding of microbial processes, including activity, growth rates, and gas conversion—is urgently needed[66]. Depleted gas reservoirs and saline aquifers are not sterile. They can host microbial life even at temperatures up to 120°C, with no strict limits on pressure or brine salinity. Indigenous or introduced anaerobic microorganisms, such as methanogens, acetogens, and sulfate-reducing bacteria, use $H_2$ as an electron donor. These microbes can convert $H_2$ to methane ($CH_4$), generate toxic $H_2S$, alter water chemistry, affect interfacial dynamics[21], change system wettability[67], and reduce injection rates[68]. While microbial issues are well-documented in hydrocarbon production, $CO_2$ sequestration, and geothermal storage, their implications for $H_2$ storage remain poorly understood. Existing knowledge is often extrapolated from shallow environments, such as soils and lakes, rather than deep subsurface systems. For example, high $H_2$ partial pressures in shallow settings have been shown to inhibit microbial growth and enzyme activity[69], but it is unclear whether similar suppression occurs in deep reservoirs.

### 2.3 Nuclear waste geological disposal (NWGD)

Nuclear energy is widely regarded as a low-carbon energy source due to its minimal direct $CO_2$ emissions during power generation. Many countries consider this to be an important way to reduce greenhouse gas emissions and achieve carbon neutrality. However, the management of nuclear waste, particularly its final disposal in deep geological repositories, presents significant scientific and engineering challenges[70-72]. These geological repositories require a multi-barrier system that includes clay-based engineered materials or natural materials known for their nano-size pores, low permeability, self-sealing properties, and chemical stability[73, 74]. Gas generation in these materials can be caused by processes such as the anaerobic corrosion of metals, the radiolysis of water, and the degradation of organic materials[73, 75]. These processes result in the formation of reactive gases like hydrogen, methane, and carbon dioxide. In the confined pore spaces of the repository, multiphase interactions between these gases, pore water, and solid materials can lead to complex phenomena such as dissolution/precipitation, sorption, or capillary-driven flows[76, 77]. If the generated gas cannot escape from nanoporous materials through diffusion, pressure build-up may induce mechanical deformation, creating fractures or pathway dilation[78]. These changes not only alter the transport dynamics but may also reduce the barrier's performance, influencing radionuclide migration. Multiphase flows inside the clay-based materials, including the interactions of gases, liquids, and reactive components, remain challenging[79]. Addressing these challenges requires a pore-scale fundamental understanding of coupled reactive and transport processes under varying thermo-hydro-mechanical-chemical (THMC) conditions[80], as shown in Fig. 1.

**Hydro-mechanical coupling.** Gas transport in porous media can switch between diffusion and two-phase flow, depending on the rate of gas generation, microstructure, and water saturation levels[78, 81]. Predicting these transitions is challenging because these dynamic interactions vary over time and depend on the evolving saturation, temperature, and stress conditions, which are difficult to replicate in core-scale experimental setups[82]. For example, while the gas pressure does not exceed the capillary entry pressure of nanopores in clays, the gas phase can create fractures or micro-pathways to enter clays, altering their permeability[83]. If these pathways do not reseal, the barrier's performance is degraded. Understanding the thresholds for such events remains difficult due to the complex hydro-mechanical coupling processes in porous media[84].

**Chemical reactions.** The interaction of reactive gases or ions, with pore water, can induce chemical reactions that modify both the flow and diffusion properties of clays[85-87]. For example, carbonate precipitation due to $CO_2$ transport can block pore spaces, changing the flow pathway[88]. Concurrently, ion transport and chemical reactions can also induce swelling or shrinkage of the clay matrix, further modifying flow pathways[89]. Geochemical feedback on transport is difficult to quantify due to the interplay of reaction kinetics, transport dynamics, and evolving system conditions. Such coupled physiochemical processes challenge the reliability of current predictive models.

## 3. Fabrication and design of microfluidic porous media

To reveal the hidden physics within opaque porous media, typically regarded as a "black box", microfluidics has been developed to visualize and simulate these complex environments. Over the past 70 years, microfluidics has evolved into an invaluable tool for investigating complex fluid flow processes in porous media (Fig.2). This technology enables researchers to design and fabricate pore structures across a wide range of length scales, including dimensions comparable to real-world porous formations. Additionally, advances in microscale detection techniques, such as high-speed imaging and cutting-edge microscopy, have significantly enhanced the ability to visualize and capture rapid flow processes with unprecedented spatial and temporal resolution, down to the micrometer, thus allowing researchers to move from *milli-fluidics* to *micro-fluidics*.

From a historical perspective, Chatenever and Calhoun[90] made the first attempt to pack a monolayer of glass beads between two glass plates in 1952, creating a transparent porous medium to investigate pore-scale mechanisms of crude oil recovery by brine displacement. The glass beads used were carefully screened to a size of approximately 178 μm and fit within a 190-μm gap between the two glass plates. To mitigate displacement by the flow of this monolayer of dust-like glass particles, a 150-mesh screen (approximately 89 μm) had to be applied at the inlet and outlet ports. Furthermore, the imaging of the liquid-liquid interface was hindered by the contours of the glass beads, complicating the image postprocessing. Later, Mattax and Kyte[91] developed a capillary network model by etching patterns onto a glass plate, coining the term "micromodel" to describe these etched glass capillary network models. They utilized this model to investigate waterflooding under both strongly and weakly water-wet conditions. The term "micromodels" has been coined to denote experimental flow cells containing model porous media consisting of grains or connected channels and allowing pore-scale visualization of multiphase flow and transport processes. Before microfluidics techniques became widespread in porous media studies, such micromodels featuring media with millimeter size pores/channels, have been widely used in subsurface engineering and dispersed media physics. With the development of the semiconductor industry, the micro/nanofabrication technique allowed the creation of small and precise complex patterns from numerical designs, at scales down to the micrometer, thus allowing researchers to move from milli-fluidics to micro-fluidics. This has been made possible by the adoption of photolithography which is a technique common to micro/nano fabrication in micro-electromechanical systems (MEMS) and the semiconductor industry. In 1979, the development of one of the first silicon-based micromodels[92] marked a significant milestone in the widespread adoption of microfluidics for exploring fluid flow in porous media, and there has since been the exploration of different materials and nanofabrication techniques as well as probe/visualization techniques to investigate different processes in porous media. In this review, the terms "microfluidic chip" and "micromodel" are used interchangeably with no conceptual distinction. While traditional microfluidics is typically defined by feature sizes smaller than 1 mm, this paper focuses on the role of microfluidics in studying fundamental multiphase reactive flow mechanisms. Therefore, some models with characteristic sizes exceeding 1 mm, such as milli-fluidic systems, are also included, provided they adhere to the principles of similitude. In Fig. 2, we show a timeline of the development of micromodels, showcasing advances in fabrication techniques, geometric designs, surface properties, and experimental detection techniques.

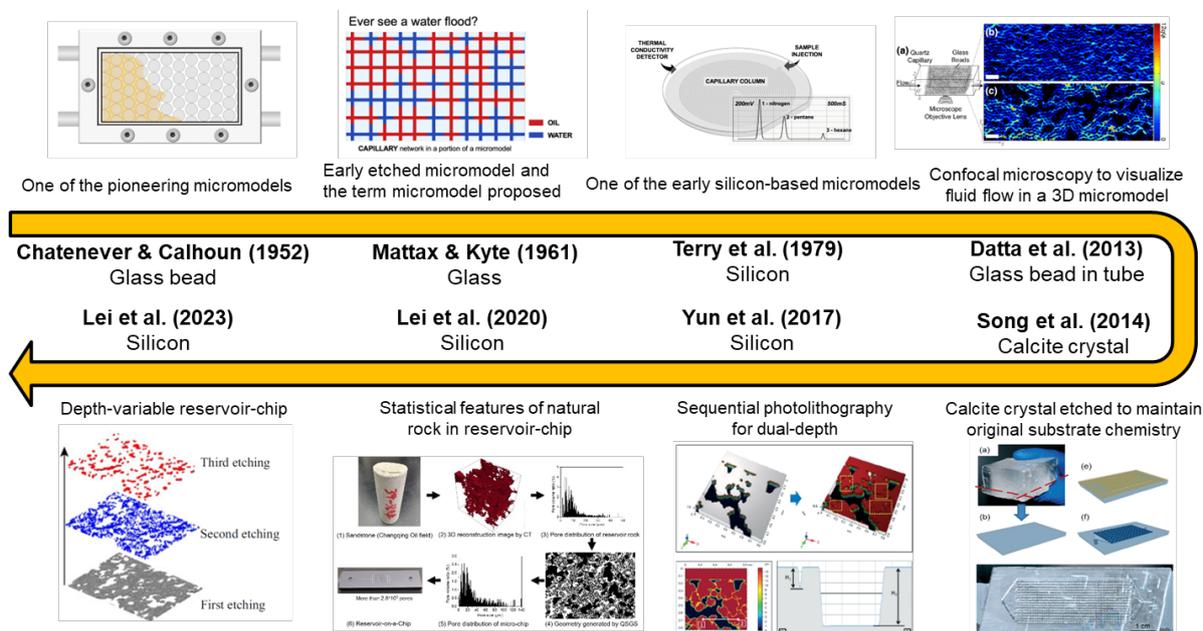

**Fig. 2.** Key advancements in microfluidic chip design over the past 70 years for mimicking porous structures in subsurface engineering. From the starting point to the endpoint are listed: The monolayer glass bead packing model (modified from Ref[90]) - one of the pioneering micromodels; the etched micromodel with the capillary network geometry (modified from Ref [91]) - the early etched glass micromodel; wet etching in silicon wafer (modified from Ref[92]) - one of the early silicon-based micromodels; visualizing flow in 3D glass bead-packed structures[93] (Adapted with permission. Copyright 2013, American Physical Society); incorporating mineral properties[94] (Adapted with permission, Copyright 2014, Royal Society of Chemistry); fabricating dual-depth of pores and throats[95] (Adapted with permission, Copyright 2017, Royal Society of Chemistry); generating reservoir chips with statistical data[96]; and developing depth-variation reservoir chips[16].

### 3.1 Materials and fabrication techniques of microfluidic porous media

To successfully utilize microfluidic porous media in research or engineering applications, it is crucial to address several key factors, including targeting the fundamental problem (multiphase and multi-physics processes), the specific environment (the pressure, temperature, surface properties, etc.), the relevant geometric features (type of porous media and characteristic scales), and the appropriate detection methods to capture the different phases and interaction processes. Typically, spatialized visualization of hidden and dynamic processes is fundamental to these microfluidic experiments. In this case, the optical properties of microfluidic porous structures are crucial. Microfluidic chips are fabricated with transparent materials including glass, silicon (transparent under infrared light[97, 98]), and polymers, such as polydimethylsiloxane (PDMS), polymethyl methacrylate (PMMA), and polytetrafluoroethylene (PTFE or called Teflon)[12, 13, 99]. Different fabrication techniques are applied depending on the material of the microfluidic substrate. Generally, the fabrication process involves creating patterns on the substrate and assembling wafers to form a functional fluid flow model. Common fabrication techniques for pattern creation include photolithography (either positive or negative), e-beam lithography, soft lithography, 3D printing, micromilling, and other advanced methods[100, 101]. Fig.3 shows how to select the cover material for the micromodel depending on the chosen detection method. Different cover materials will be bonded to a patterned compatible substrate, where the patterning method also depends on the material. For example, a classical micromodel using silicon substrate can be fabricated using photolithography followed by dry etching with plasma to create a porous media pattern. The substrate is then bonded to a glass cover through anodic bonding, forming a sealed microfluidic structure.

**Lithographic techniques and corresponding materials.** In subsurface engineering, the hardness and stiffness of rock under high-pressure, high-temperature conditions necessitate the use of materials such as glass and silicon to fabricate microfluidic porous media. These materials not only offer precise geometry control but also exhibit robust chemical stability, making them the most widely used option in such environments[100]. Sapphire is also considered a viable option due to its extreme hardness, second only to diamond and moissanite, as well as its high stiffness and chemical stability. However, sapphire has seen limited use in microfluidic devices because pattern transfer and bonding during fabrication are challenging[102]. Depending on the research goal, such as delineating a multiphase flow process in porous media that mimics unconsolidated sandstone, the core step is transferring a pattern onto the substrate. Although wet etching is also widely used, it typically has a slower etch rate and produces isotropic profiles, which limit vertical sidewall control and make deep etching difficult. It is also important to note the differences between wet etching in silicon and glass. In glass wafers, etching is inherently isotropic, resulting in rounded channel walls unless specifically controlled, as $SiO_2$ lacks crystallographic orientation. In contrast, wet etching of silicon is anisotropic, typically producing 54.7° sidewalls due to its crystal structure. This occurs because the <100> plane has the lowest atomic density, leading to a faster etching rate, whereas the <111> plane has the highest atomic density, making it more resistant to etching[103].

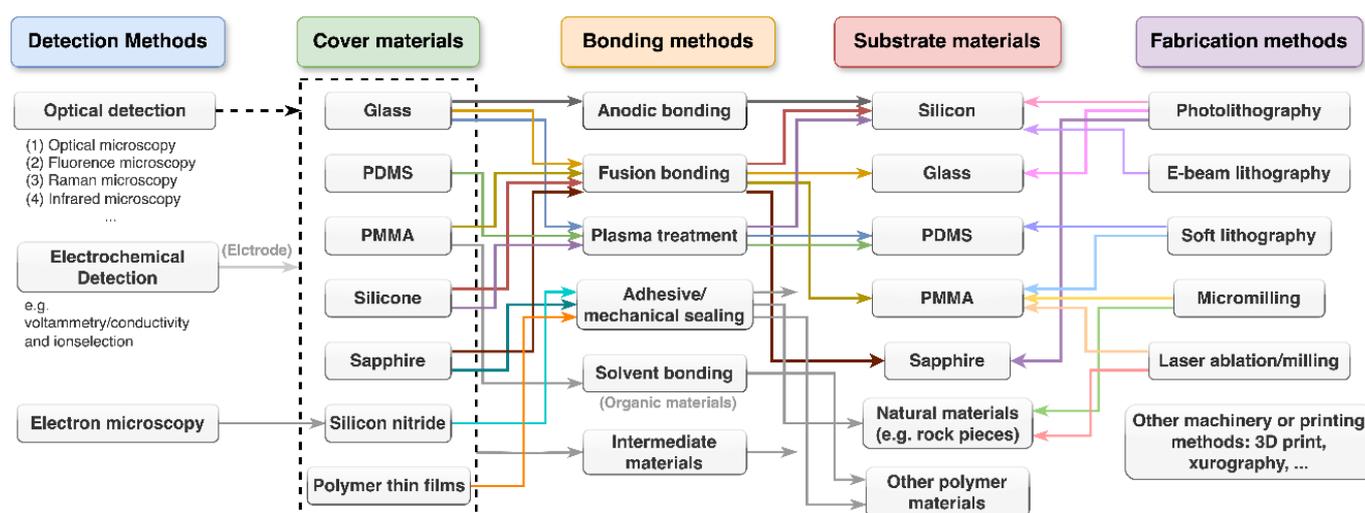

Fig. 3. A summary of materials used in microfluidic fabrication. The selection of cover and substrate materials is based on their properties, which influence bonding methods, detection techniques, and fabrication approaches. Optical detection is most used for microfluidics. In addition to photolithography, recent advancements in fabrication techniques for different materials are highlighted. Grey-colored links indicate connections to broader kinds of materials.

We will use a silicon wafer as an example to briefly outline a typical photolithography process for microfluidic fabrication[104]. Before conducting photolithography, a clean silicon wafer is typically treated with hexamethyldisilazane (HMDS) vapor to enhance photoresist adhesion. A thin layer of photoresist is then applied to the wafer surface, serving as a protective mask to define specific regions shielded from the subsequent etching process. Depending on the desired etching depth, various photoresists (with different viscosities) can be spin-coated onto the wafer to achieve the required thickness. In later photolithography steps, especially when covering previously etched patterns, spray coating may be used instead of spin coating. After photoresist coating, a soft bake is performed to remove solvents from the photoresist and improve film uniformity. Next comes the exposure step: a photomask is placed over the wafer, and UV light shines through the transparent areas of the mask. With a positive photoresist, the exposed regions become soluble in the developer, so they can be removed in the subsequent development step, leaving the unexposed areas to protect the underlying substrate during the next etching step. If using negative photoresists (e.g., SU-8, commonly used for soft lithography molds), the exposed regions crosslink and harden under UV light, so the unexposed regions can be removed during the development process. Nowadays, for rapid prototyping, maskless aligners are also commonly used, allowing users to write custom patterns without the need for a physical photomask. After exposure, a development process is to remove the photoresist from specific regions, and a hard bake process will follow to finalize the pattern indicated by the photoresist. Then, the wafer is ready for the final etching step, which permanently transfers the pattern onto the substrate. Two types of etching approaches, dry and wet etching, are typically used for pattern creation. In dry etching, a plasma or reactive-ion environment bombards the wafer surface with energetic ions, breaking molecular bonds and selectively removing material through chemical reactions or physical sputtering[105]. In the wet etching process, the wafer is immersed in a reactive solution, typically hydrofluoric acid (HF) or buffered oxide etch (BOE) solutions, for a duration determined by pre-measured etch rates. The acid selectively dissolves the unprotected regions where the photoresist is absent, allowing precise control over the etch depth. Vapor etching, which utilizes HF vapor to etch the wafer, is an alternative method primarily used for structures that may be damaged during the rinse and drying process. Similar to photolithography, electron-beam lithography (E-beam lithography, EBL) is a widely used technique for creating nanoscale patterns, as the wavelength of UV light in traditional photolithography cannot achieve sub-10 nm resolution[106]. The key exposure step of EBL is the same as scanning electron microscopy (SEM), where an electron beam gun emits focused electron beams in a high-vacuum chamber onto the surface of a substrate. By prolonging the exposure, the electron beam can locally burn the material, enabling precise pattern creation. During EBL, electrons continuously bombard specific regions of a wafer coated with electron-sensitive resist, defining the desired pattern. Following exposure, the process proceeds with development and etching, similar to photolithography. Soft lithography involves fabricating a master pattern on a substrate, using it as a mold, and then pouring a polymer mixture onto this mold[107]. Once cured, the patterned soft material is peeled off, replicating the mold's features. A classic example is PDMS microfluidics, which excels in sealing diverse materials and bio-related applications.

**Non-lithographic techniques.** Due to the complexity of lithographic techniques, recent research has explored alternative microfabrication approaches, including 3D printing and micro-milling. Advances in 3D printing have significantly improved the precision of microfluidic fabrication. In addition to additive manufacturing techniques that use nozzle-based material extrusion or jetting, recent developments have focused on stereolithography (SLA), where photosensitive polymers are solidified using focused optical energy, such as laser exposure[108, 109]. Another emerging approach is direct-write lithography, in which the photoresist is directly patterned to create microstructures. This technique enables the fabrication of 3D microstructures by varying the focus depth during exposure. Cheng et al. [110] demonstrated a method using deposited photoresist to create microstructures sealed between two glass plates, achieving an ultrathin structure (~1 µm thick). However, these approaches have certain limitations. For example, the flow characteristics in microfluidic devices fabricated using photoresist-based methods can be influenced by surface properties, and photoresist materials are generally susceptible to dissolution in organic solvents. In this case, oil-based and organic fluids are often incompatible with these micromodels, restricting their applicability in certain experiments.

Micromilling has emerged as an alternative promising technique for microfluidic fabrication, as it physically sculpts designed patterns onto substrates through computer numerical control (CNC) without requiring complex or time-consuming processes. This method is considered an efficient and rapid prototyping approach for developing microfluidic devices[111, 112]. Bao et al.[113] highlighted micromilling and laser ablation as potential fabrication techniques for microfluidics in subsurface engineering, particularly for commercial applications. Micromilling can achieve about 50 µm feature sizes on metal substrates, while laser cutting has been demonstrated on silicon wafers. Although its resolution is lower than lithographic methods, it remains suitable for specific applications. However, its large-scale adoption in subsurface engineering microfluidics is still limited. Some studies have used micromilling for mold fabrication in soft lithography[114, 115]. For example, Gao et al.[116] developed a low-cost micromilled micromodel by milling a rectangular array on pyrophyllite, producing trapezoidal channels (250 μm top, 200 μm bottom) at a fabrication cost under $10. Despite its advantages, micromilling faces key challenges in microfluidic porous media fabrication. Since it is a direct-write method, every feature must be individually milled, making complex porous patterns time-consuming and increasing tool wear. Precision declines over extended operations, and achieving micron-scale features remains difficult. Surface roughness control is another limitation, as tool path artifacts and material adhesion can reduce feature definition. While micromilling offers rapid and flexible prototyping, its limitations in resolution and surface quality must be addressed before it can be widely applied in microfluidic porous media research.

**Biological microfluidic chips.** Microfluidic chips are widely utilized in microbial research as they offer precise control over microenvironmental conditions and support high-resolution imaging. Thus, they enable the study of microbial behavior at the scales of both individual bacteria and bacterial population, where the same scale ratio between confining medium and bacterial size as in real conditions can be achieved in laboratory experiments (for example, subsurface porous media or industrial reactors) . The design and processing of microfluidic chips for microbe-related research requires special considerations compared to ordinary non-biological microfluidic chips:

Sterility: The choice of materials must align with the required sterilization protocols. Biological chips must undergo sterilization without compromising material integrity or function. Common sterilization methods include UV irradiation, ethanol rinsing, and autoclaving (for heat-resistant materials).

Gas Exchange: Maintaining an adequate supply of oxygen or carbon dioxide for microbial growth is crucial (depending on whether the bacteria are oxic or anoxic). Gas-permeable materials such as PDMS are beneficial in this regard.

Nutrient Delivery: Continuous or gradient-based delivery of nutrients necessitates designs that prevent clogging. Therefore, chip materials also need to be compatible with the chemical composition of the nutrients contained.

Real-Time Monitoring: Optical clarity and compatibility with imaging systems are essential for tracking microbial growth, displacement, attachment to surface and biofilm formation, and general behavior.

Surface inertness: Biological molecules such as proteins, peptides, and nucleic acids interact with the chip's surfaces, resulting in adsorption, enzymatic activity loss, or denaturation, which can compromise experimental results[117, 118]. Strategies to mitigate these interactions include surface modifications. Another suitable biomolecule is added in excess to block the substrates, such as passivation with bovine serum albumin (BSA)[119] or polyethylene glycol (PEG)[120].

Moreover, since microbial activity visualization and quantification often rely on fluorescence or laser-based technologies, biological microfluidic chips must not only be transparent but also exhibit low autofluorescence. Given that bacterial tracking and observation can extend over weeks or even months, chip materials must remain stable under prolonged exposure to high-energy light sources.

**Material comparison and selection.** Optical methods are commonly used in microfluidic experiments, requiring an optically transparent cover for most device designs. While silicon is opaque to visible light (but transparent under infrared[97]), it is often bonded to glass to enable optical visualization. Different bonding techniques are used depending on the substrate and cover materials. For example, anodic bonding is commonly used for silicon-glass interfaces, where heat and voltage drive alkali ion migration in glass, creating an electrostatic attraction that forms a strong, permanent seal. Material surface properties also vary. Silica surfaces are hydrophilic with high charge density and well-characterized chemistry, while polymer surfaces are typically hydrophobic with lower and sometimes unpredictable surface charge[121, 122]. Glass, one of the earliest materials used in microfluidic porous media research, remains a reliable option, particularly when surface properties are not a primary concern. Polymer-based micromodels, such as those made from PDMS and PMMA, offer cost-effective and simpler fabrication. PDMS, with its porous Si-O matrix, is gas-permeable (e.g., to oxygen and $CO_2$), making it ideal for biological and subsurface applications. However, it swells in nonpolar solvents (e.g., hydrocarbons, toluene) and has limited chemical, mechanical, and thermal stability, restricting its use in organic solvents-related studies[123]. For biological microfluidic chips, PDMS facilitates microbial growth by ensuring appropriate oxygen exchange. PDMS also exhibits poor adhesion to mammalian cells, whereas alternative materials like PFPE (perfluoropolyether) offer improved chemical inertness. However, they are less permeable to gases, restricting their use to a subset of biological studies. Studies have shown that biological cells maintain normal viability when cultured in Teflon channels[99]. Compared to PDMS, whole-Teflon chips offer several advantages, including the absence of small molecule absorption, minimal biomolecule adsorption to channel walls, and no leaching of residual compounds from the material into the solution. PMMA provides better chemical resistance and mechanical stiffness than PDMS but lacks the durability and precision of glass. It is well-suited for rapid prototyping since micromilling or laser ablation can create ~100 μm features with ±2 μm tolerance without extensive processing[112]. Off-Stoichiometric Thiol-Ene (OSTE) polymers offer the mechanical and chemical properties of both PDMS and commercial thermoplastics[124], with easy surface modifications and UV-based bonding[125]. Norland Optical Adhesive (NOA), a light-curing thiol-based resin, is also a promising material for microfluidic chips, in particular for studies involving bacteria; microfluidic cells entirely made of NOA have also been designed to study the dissolution trapping of $CO_2$ [46]. It is known for its excellent optical transparency, low autofluorescence, and resistance to a wide range of organic solvents. Used in microfluidic fabrication via "microfluidic sticker" soft lithography[126], NOA enables bonding with glass or silicon and allows device recycling by dissolving components in chlorinated solvents[127]. NOA is also impermeable to gases and water vapor, with high elasticity to prevent channel deformation under high-pressure flow, though thin NOA layers retain some flexibility for easier bonding[128]. Moreover, the properties of NOA are complementary to those of PDMS, so it is possible to consider using both NOA and PDMS to fabricate composite microfluidic chips. NOA can be used in channels requiring high-pressure flow or exposure to organic solvents, while PDMS can serve areas that house oxygen-exchanging microbial, the oxygen being provided through molecular diffusion across the PDMS layer from another gas-impermeable channel where oxygen is being circulated[129]. It is worth noting that NOA does not adhere directly to PDMS, necessitating the use of a plasma device and an intermediate glass layer for bonding[128].

### 3.2 Geometrical design of microfluidic porous media

The geometrical design of porous structures has been a pivotal focus in microfluidic experiments, enabling researchers to replicate the intricate geometries of natural porous media. Over the past two decades, significant efforts have been made to incorporate the statistical characteristics of natural porous media into microfluidic systems. Early approaches involved creating simplified replicas by using 2D slices of real-world 3D porous media as templates for microfluidic designs[130]. While these slices captured certain geometric features, they failed to represent the full statistical complexity of natural rock structures. Additionally, the unconnected nature of many 2D slices required artificial modifications to ensure functionality. A major breakthrough was achieved by Kumar Gunda et al.[131], who introduced a method to integrate statistical network information from natural rocks into microfluidic designs. Using stochastic random network generators and Delaunay triangulation, they conceptualized the "reservoir-on-a-chip." While this approach represented a significant advancement, it fell short of fully replicating critical statistical properties such as pore size distribution and hierarchical structural features. Subsequent innovations have addressed these limitations. Lei et al.[96] employed the Quartet Structure Generation Set (QSGS) to regenerate 2D porous structures on microfluidic chips, accurately preserving the pore size distribution and structural features derived from 3D micro-CT images of natural rocks. Building on this design, additional structural information, such as preferential flow pathways[132, 133], has further enhanced the realism of reservoir-chip models. However, they are still limited by the geometric confinement from the uniform depth in microfluidic chips.

To consider 3D geometrical features, randomly packing particles between transparent parallel plates [134-136] or advanced 3D printed models[137, 138] have become important for studying displacement mechanisms. Advanced imaging and reconstruction methods, including confocal microscopy, light-induced fluorescence (LIF) scanning[139], and X-ray computed tomography, enable extension into three dimensions, allowing for the realization of truly 3D geometrical features of natural porous media. However, these visualization methods face inherent limitations in studying flow dynamics, particularly in terms of structural accuracy, repeatability, multiphase fluid selection, and synchronous imaging of the full flow field during displacement processes. To balance 3D pore structure with fast optical visualization of flow dynamics, hybrid models such as 2.5D microfluidic porous media have been developed. These models feature 3D geometric characteristics in the pore space while maintaining a planar network topology, similar to 2D models. For example, a simple glass isotropic etching method was applied to create square arrays connected by shallower throats in microfluidic chips, promoting unstable snap-off phenomena compared to traditional chips with uniform depth[140]. In silicon microfluidic chips, sequential photolithography and multi-etching techniques have enabled the creation of pores and throats with varying depths but within the porous geometry of 2D slices of natural porous media[95]. These innovations have revealed distinct flow behaviors in single- and dual-depth micromodels during both imbibition and drainage stages. Recently, Lei et al.[16, 141] combined statistical reservoir data with sequential photolithography and multiple etching to create a 2.5D reservoir-chip, integrating 3D structural information with the advantages of microfluidic systems, as shown in Fig. 4A. Compared with traditional 2D microfluidic porous media with uniform depth, this 2.5D microfluidic porous media can release the strong hydraulic diameter limitation and present the 3D pore structure as the natural engineered porous media (Fig. 4B).

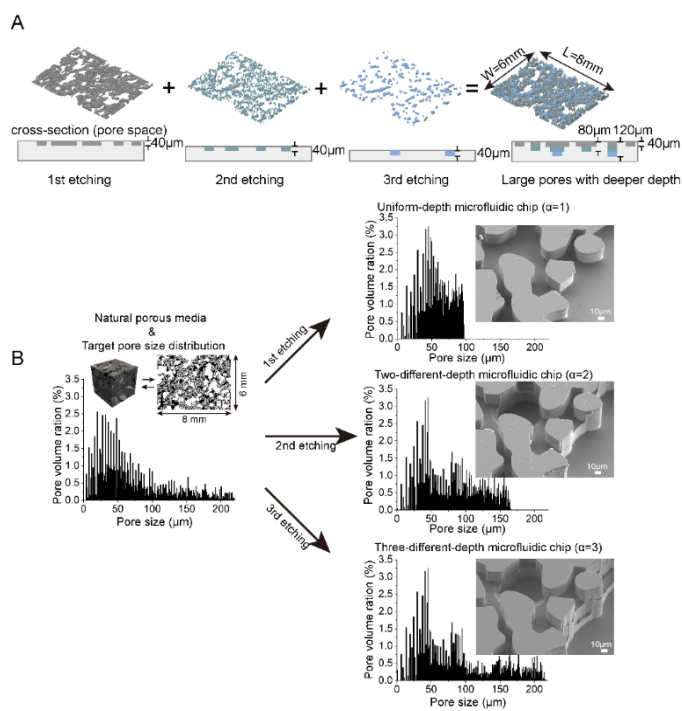

**Fig. 4.** Microfluidic chips with different depth variations to mimic 3D porous media. (A) Schematic showing the strategy for the fabrication of depth-variable microfluidic chips by optimized sequential photolithography and multiple etchings[16]. (B) Comparing pore size distributions in natural-occurring 3D porous media and microfluidic chips[141].

### 3.3 Surface properties modification of microfluidic porous media

To mimic the surface properties of natural porous media, different functional properties are incorporated into microfluidic chips, as shown in Fig. 5. Wettability, a key parameter describing the affinity of invading or defending fluids to solid surfaces, is one of the most critical factors influencing flow in porous media. For homogeneous wettability alterations, different microfluidic materials exhibit varying wettability characteristics due to differences in surface energy. Even among glass substrates, wettability differences are evident: Schott BF33 is more hydrophilic than Schott B270, offering diverse options for experimentation. Furthermore, wettability can be tailored using techniques such as thermal oxidation[142], silanization[143], or surface coatings[144], or oxygen plasma exposure. In particular, the latter method allows tuning the wettability of NOA continuously from 0 to 110°, i.e. over the entire hydrophilic range and 20° of the weakly-hydrophobic range[145]. For heterogeneous wettability, mixed wettability within porous media can be achieved by leveraging flow inhomogeneity during cyclic injections of coating and immiscible fluids[146, 147]. However, this approach is often constrained by challenges in controllability and repeatability due to complex multiphase flow dynamics. Recent advancements in combining sequential photolithography and thin-film deposition have enabled precise, localized surface wettability modifications in microfluidic porous media. For instance, molecular vapor deposition of perfluorodecyltrichlorosilane selectively alters the wettability of specific substrate regions[148, 149], creating hydrophobic coated areas while retaining the original wettability in non-coated regions, thus offering highly localized control. Additionally, natural mixed wettability can be mimicked by creating hydrophilic and superhydrophobic micropatterns simultaneously through area-selective self-assembly of monomers[150].

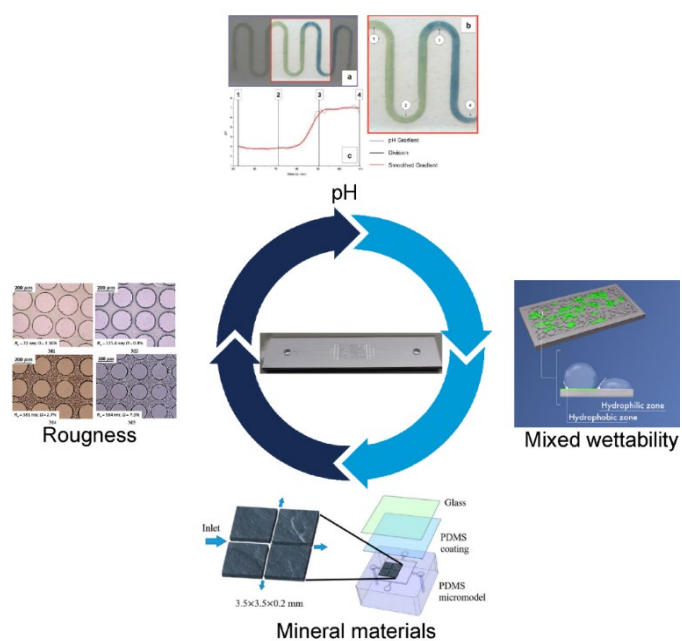

**Fig.5.** Functionalization of microfluidic porous media by adding roughness[151] (Reproduced with permission, Copyright 2019, American Geophysical Union), mineral properties[19] (Reproduced with permission from the author), mixed wettability[148] (Reproduced with permission under a CC-BY Creative Commons license, Copyright 2023, Royal Society of Chemistry), pH visualization[152] (Reproduced with permission, Copyright 2013, Royal Society of Chemistry).

Similarly, based on polyaniline coatings, real-time, reagent-free pH monitoring along entire channel lengths as pH optical sensing can be integrated into microfluidic devices. These devices feature low-cost optical detection systems for spatial mapping of pH gradients via digital imaging, providing a versatile platform for dynamic chemical sensing[152, 153]. To modify surface roughness, glass etching cream—composed of fluoride salts (e.g., sodium fluoride or ammonium bifluoride), thickeners (e.g., polysaccharides, hectorite clays, and polyurethanes), organic solvents (e.g., dipropylene glycol methyl ether acetate), and water—can be used to generate roughness in glass micromodels[151, 154]. Variations in roughness on soda-lime glass substrates can be achieved by diluting the etchant with deionized water.

In recent years, geomaterial microfluidic chips have gained attention for their ability to mimic the chemical and physical properties of natural porous media. Direct etching of pore structures into natural crystals allows for the study of preferential dissolution processes influenced by flow fields and crystallographic orientations[94]. Surface structure information can be incorporated to better understand displacement effects. For example, joint-fracture microfluidic channels designed on natural coal samples demonstrate that fracture roughness results in higher pressure differences compared to traditional PDMS chips[155]. Using original minerals for research on liquid-solid chemical reactions provides a more direct approach to studying the mechanisms of multiphase reaction flow[19]. Simple channels etched into natural crystals enable detailed exploration of these multiphase reaction flow processes[156]. Additionally, integrating mineral grains into microchannels benefits the direct observation of calcite dissolution processes[157], some studies even incorporate the small mineral grains into traditional micro-nano fabrication materials, such as PDMS, allowing simultaneous study of flow in pore structures and chemical reaction kinetics of solid walls[158].

### 3.4 Recent advances and future development

Recent advances across the field of microfluidic porous media, which encompasses micro-nano processing technology, fluid mechanics, and sustainable energy, have been made possible by unprecedented innovations within microfluidic design and fabrication.

**Towards nano and multiscale**. Recent advancements in microfluidic research have begun to focus on nanoscale phenomena. Given the multiscale nature of natural porous media and the increasing interest in flow phenomena within ultra-low permeability porous media, nanostructures are becoming a focal point of microfluidic studies. In nanoscale porous structures, capillary forces are significantly amplified, phase condensation behaviors become pronounced, and electrokinetic effects become dominant. A notable development in this area is the emergence of "nanofluidics", which has been proposed as a solution for sustainable energy applications in ultra-low permeability porous media. This topic has garnered significant attention recently[159]. For instance, in low-permeability porous media such as shale, typical pore sizes are below 100 nm, with porosities often under 10%. Although efforts have been made to explore nanoscale processes in porous media, challenges remain in both the fabrication and application of nanoscale microfluidic systems[160, 161]. One key challenge lies in achieving the extreme precision required for nanoscale fabrication. Traditional photolithography methods are insufficient for creating features on the scale of hundreds of nanometers due to the limitations of ultraviolet light during the exposure process. Instead, electron-beam lithography (e-beam lithography) is typically employed to reliably produce nanoscale features. As a result, most current research has focused on planar channels, where the width is on the micron scale, but the depth is reduced to the nanoscale, enabling at least one dimension to reach the nanoscale regime. Moreover, as feature sizes in porous media scale down, designing patterns with numerous pores becomes increasingly complex. Consequently, many studies use single straight channels for experimentation instead of replicating complex porous media patterns[160]. Another fabrication issue is ensuring surface smoothness and preventing dust contamination, as these imperfections can compromise the accuracy of the nanofluidic models. Bonding in nanofluidics presents further challenges. For example, it was demonstrated that the aspect ratio of channels must be carefully controlled during bonding to avoid the collapse of the top cover onto the channel bottom, which can seal the void space unintentionally[162]. Visualization techniques also encounter significant hurdles. At the nanoscale, dimensions fall below the wavelength of visible light, making it difficult to differentiate between the light reflected from the top cover and the bottom substrate of a nanofluidic channel. To address this, a method using silicon nitride $Si_3N_4$ deposition was developed to enhance fluorescence intensity, enabling optical detection through a microscope even in sub-10 nm channels[160]. In nanofluidics, future research on the coupling effect and heterogeneity between nano features and microscale features in porous media should be further explored. However, the coupled scale will further complicate the fabrication process.

**Electronics and multi-physical detection methods.** Typically, microfluidic channels are sealed between a substrate and a cover, making contact-based imaging methods, such as SEM impractical. However, a recent study demonstrated a novel approach using a $Si_3N_4$ film that allows electrons to penetrate the membrane, enabling SEM imaging of fluid flow within nanoporous structures[163]. Silicon nitride membranes are commercially available, but they generally come in millimeter-scale frames and are relatively thin, meaning they can deform under pressure and are unsuitable for high-pressure injections. Despite these limitations, integrating $Si_3N_4$ membranes provides new avenues for SEM-based detection in microfluidics. Also, when combined with conventional detection methods, it can help quantify mineralogy in reactive transport processes. Recently, a PDMS-based microfluidic chip for geoelectrical monitoring of calcite dissolution processes in a linear channel was developed by integrating electrodes directly into the device[164, 165]. They employed a spectral induced polarization (SIP) method to assess conduction in porous media and successfully obtain electrical characterization signals of the geochemical reactivity on a microfluidic chip. This setup enabled real-time tracking of complex electrical conductivity changes during the reactive transport of calcite ($CaCO_3$) dissolved by hydrochloric acid (HCl)[164] or precipitation by injecting two reactive solutions ($CaCl_2$ solution and $Na_2CO_3$ solution)[165]. However, calibration proved challenging due to numerous variables, including the microfluidic structure and the specific physicochemical processes[166, 167]. Moreover, fabricating embedded electrodes within the microfluidic channels is relatively complex, though the authors used PDMS to streamline the process. Note that earlier studies used NOA micromodels also including electrodes for the characterization of solute transport and calcite precipitation through real[172] and complex[165] conductivity measurements in disordered 2D porous media with mm-scale pores. If considering the pressure conditions and using a silicon substrate, future directions might involve direct electrode deposition or leveraging the semiconductor properties of the silicon substrate to simplify electrode integration and enhance detection capabilities. In Table 1, we make a list of detection methods that have been utilized in microfluidic porous media research which could serve as a reference for future explorations.

Table 1. Detection methods utilized in microfluidic porous media research

|  | Objects | Pros | Cons | Ref. |
|---|---|---|---|---|
| Optical microscopy | Solids and fluids | Adjustable field of view and quick visualization with high-speed camera | Transparent object or observation of the outer surface | 16, 133, 141 |
| Laser confocal miscopy | Solids and fluids | High resolution and 3D imaging | Transparent object or observation of the outer surface | 93, 135, 136 |
| Raman spectroscopy | Solids and fluids | Non-destructive detection of materials and small sample volumes | Long acquisition time and background noise | 168 |
| SEM | Solids | Observation of the nanoscale | No real-time capture; small field of view; conductive or with coater | 163 |
| X-ray | Solids and fluids | 3D pattern; different resolution, mm, micro, nano-CT | Time of scanning; real-time in development | 169 |
| NMR/MRI | Specified fluids | 1D composition or 3D spatial distribution | Low spatial resolution (~0.1 mm) | 170, 171 |

**Geomechanically functional microfluidics.** Most microfluidics research focuses on visualizing fluid flow and probing fluid behavior, with relatively little attention paid to the coupled effects of fluid flow and geomechanics. Some microfluidic chips with simple geometries, such as Hele-Shaw models – two plates separated by a small, uniform gap - have been fabricated using thin elastic sheets like polypropylene or latex to investigate the effects of solid deformation on interfacial phenomena[15, 173, 174]. Although these studies primarily focus on physiological networks, such as microvasculature, organs-on-a-chip, and the hierarchical lung airway network, they provide valuable insights into fluid-structure interactions. Recently, a 10 cm-diameter cylinder made of stereolithographically 3D-printed, optically clear PMMA was used, with fluid injected through a central hole. This setup enables the formation of an uninterrupted, extended fracture front, with the fluid controlling the loading conditions that determine the amplitude of the forward jump[175]. However, no published studies have yet explored microfluidic porous media in this context, making it a promising direction for future research. Brittle substrates like silicon or glass are generally unsuitable for fluid-solid coupling because such applications require materials with expandable or compliant properties. While PDMS could potentially accommodate mechanical deformation, it cannot withstand the high pressures often encountered in geological settings. Additionally, external pressure must be applied in a way that still allows optical access for monitoring fluid conditions. In this case, a single axial pressure is easier to apply than three axials for microfluidic chips. Despite these challenges to mimic in-suit geomechanics conditions, we believe that time is ripe to develop fluid-solid coupled microfluidic porous media.

## 4. Physics of multiphase reactive flow in porous media

Advances in microfluidics for sustainable energy have focused on understanding the complex dynamics of multiphase reactive flows at the pore scale. These nonlinear processes involve interactions between multiphase flow, mass transport, and biogeochemical or microbially induced reactions. In this section, we will first introduce the basic governing equations and dimensionless parameters of multiphase reactive flows, and then present the microscopic mechanism of multiphase flow in porous media, as well as the effects of chemical reactions and microorganisms. We consider these processes at the *hydrodynamic scale*, that is, the scale at which fluid particles are defined and the equations of fluid mechanics are expressed; in porous media science, this is also referred to as the pore scale. This means that, in porous structures, we consider the flow within the pore space, between the geometrically complex solid structures.

### 4.1 Theoretical background on pore-scale multiphase reactive flow

Multiphase flow in porous media is not only influenced by inertial, viscous, as well as body forces, such as gravity ($f_b=g$), but also by capillary forces which act only at fluid-fluid interfaces. Inertial forces arise from the convective acceleration of fluids, while viscous forces stem from molecular momentum and contribute to energy dissipation. For incompressible Newtonian fluids, such as water, for which viscous stresses depend linearly on the local strain, the incompressible Navier-Stokers (N-S) equations describe the flow through the conservation of mass and momentum:

$$\nabla \cdot \boldsymbol{u} = 0 \quad (1)$$

$$\frac{\partial (\rho \boldsymbol{u})}{\partial t} + \nabla \cdot (\rho \boldsymbol{u} \boldsymbol{u}) = -\nabla p + \mu \nabla^2 \boldsymbol{u} + \boldsymbol{f}_b \quad (2)$$

where $t$ denotes time, $\rho$ is the fluid's density, $\mu$ is its dynamic viscosity, and $\boldsymbol{u}$ is the flow velocity.

**Capillary Pressure and Interface Dynamics.** During immiscible multiphase flow processes, surface tension, acting only at fluid-fluid interfaces due to the fact that molecules always prefer to be surrounded by identical molecules of a different nature, induce pressure discontinuity across the interfaces. It is usually denoted to capillary pressure $P_c$. For a fluid-fluid interface in thermodynamical equilibrium, $P_c$ is related to the interface tension $\gamma$ and principal radii of curvature $r_1$ and $r_2$ by the Young-Laplace equation:

$$P_c = P_{non-wetting} - P_{wetting} = \gamma \left(\frac{1}{r_1} + \frac{1}{r_2}\right) \quad (3)$$

**Mass Transport and Surface Reactions.** Solute transport in fluid phases is described by the advection-diffusion equation:

$$\frac{\partial c}{\partial t} + \boldsymbol{u} \nabla c = D \nabla^2 c + S \quad (4)$$

where $c$ represents the concentration of solute, $D$ the molecular diffusivity of the solute, and $S$ a source or sink term which can result from the fact that the solute species is a reactant or a product in a chemical reaction occurring in the fluid bulk (i.e., a heterogeneous reaction). If the solutes carry charges (i.e. ions), the above equation fails to consider Coulombic effects of ions that maintain the local electroneutrality[176, 177]. To keep the charge balance, several recent studies employed Nernst-Planck-based equations to describe solute transport in porous[73, 178, 179] and fractured media[167]. For heterogeneous reactions (i.e., involving the solid phase), the mass exchange at the fluid-solid interface is expressed by the constraint that the increase rate of the surface concentration on the wall, $\Gamma$, must be equal to the incoming diffusive solute flux:

$$-D\frac{\partial c}{\partial n} = \frac{\partial \Gamma}{\partial t} \quad (5)$$

where $n$ denotes the space coordinate in the outward normal direction of the solid wall. Adsorption and desorption are assumed to be described by first-order reactions:

$$\frac{\partial \Gamma}{\partial t} = k_a c - k_d \Gamma \quad (6)$$

where $k_a$ and $k_d$ are the dimensional adsorption and desorption rate constants, respectively. At equilibrium, the surface concentration is linearly proportional to the solute concentration, $\Gamma = k_a/k_d \, c$, which is referred to as a linear isotherm.

**Characteristic Dimensionless Numbers.** The above transfer processes in porous media are characterized by the following dimensionless numbers:

The Bond number (Bd) is an estimate of the ratio of the typical magnitude of the gravitational force to that of capillary forces:

$$Bd = \frac{\rho g r^2}{\gamma} \quad (7)$$

where $r$ is the characteristic pore size. When $Bd \ll 1$, the gravitational force can be neglected, a common assumption in microfluidic research when flow cells are positioned horizontally.

The capillary number (Ca) is the typical ratio of the magnitude of viscous forces to that of capillary forces:

$$Ca = \frac{\mu U}{\gamma} \quad (8)$$

where $U$ is the characteristic flow velocity. Note that other formulations exist[180], involving the permeability of the medium and the squared mean pore size $r^2$.

Viscosity ratio (M) is the ratio of the invading fluid's viscosity $\mu_i$ to that of the defending fluid, $\mu_d$:

$$M = \frac{\mu_i}{\mu_d} \quad (9)$$

Péclet number (Pe) quantifies the relative magnitudes of advective and diffusive transport rates.

$$Pe = \frac{rU}{D} \quad (10)$$

Note that, even in the dynamic multiphase flow conditions induced by $CO_2$ injection, a large portion of the reactivity of target storage formations may be located in low porosity and permeability material, where molecular diffusion is the dominant transport mechanism.

Damköhler numbers ($Da_I$ and $Da_{II}$) quantify the typical ratio of the characteristic time for transport ($t_A$, or $t_D$, depending on whether one considers advective or diffusive transport), to the typical reaction time $t_R$:

$$Da_I = \frac{t_A}{t_R} = \frac{rk}{U} \quad (11)$$

$$Da_{II} = \frac{t_D}{t_R} = \frac{r^2 k}{D} \quad (12)$$

where $k$ is the reaction rate constant.

**The principle of similitude.** Except for the viscosity ratio, the above characteristic non-dimensional numbers arise naturally when non-dimensionalizing the flow and reactive transport, as sole non-unit coefficients to various terms in the equations, thus rendering some of the terms dominant or negligible with respect to others, depending on whether the non-dimensional number is large or small. This explains why a process observed in the lab or in a numerical simulation is only relevant to the real system (natural or engineered) if the relevant characteristic non-dimensional numbers are the same in both. In particular, two-phase flow and reactive transport studies without bacteria. It is not mandatory for the experimental micromodel to be at the scale of the real porous medium, provided that the *M*, *Ca*, *Pe*, and *Da* are in a range that is relevant to the real system. In fact, before the birth of microfluidics, many seminal results on two-phase in disordered porous micromodels (see section 4.2 below), had been obtained on micromodels with typical pore scale in the mm range and decimetric to metric dimension, hence *millifluidic* systems[181-184], and nowadays both microfluidic and millifluidic models are used for such studies.

### 4.2 Multiphase flow dynamics in porous media

Characterizing interfacial phenomena at the pore scale using micromodel experiments is fundamental to understanding multiphase flow in porous media, a critical aspect of evaluating feasibility and managing risks in the sustainable energy industry.

**I. Confined multiphase flow dynamics in structured microchannels**

Developing constitutive models for these phenomena requires investigating flow dynamics in structured microchannels. These structured microchannels, derived from porous structures, comprise interconnected pores and throat channels with varying depths, aspect ratios, pore-to-throat size ratios, coordination numbers, and other geometric characteristics. Observation and quantification of interfacial phenomena using microfluidic experiments and the building of corresponding constitutive models forms the cornerstone of physical understanding and upscaling strategies for industrial applications involving porous media.

**Hele-Shaw model.** The Hele-Shaw model describes fluid flow confined between two closely spaced, parallel plates, where the gap between the plates is much smaller than the in-plane dimensions, creating a quasi-two-dimensional geometry. This model was initially proposed more than 125 years ago as an analog of the two-dimensional porous medium, since the flow velocity averaged over the gap is proportional to the gradient of the pressure field, which means that depth-averaged flow is controlled by Darcy's law[185]. Since then, it has proved pivotal for studying viscous flows, as well as understanding interfacial instabilities and complex pattern formations in confined spaces[186, 187]. One of the most iconic phenomena observed in Hele-Shaw cells is viscous fingering, which occurs during the displacement of a more viscous fluid by a less viscous one, as shown in Fig. 6A. This process generates nonlinear, finger-like patterns at the interface, emblematic of systems with nonlinear dynamics. The geometry of the Hele-Shaw flow channel is typically characterized by the large aspect ratio $\alpha = W/H \gg 1$ (where $H$ is the depth and $W$ is the width), which presents unique flow dynamics. In the absence of gravity and inertia, the interface becomes unstable to wavelengths greater than the critical length scale of viscous fingers $\lambda_c^* = \pi W (3\alpha^2 Ca(1-M))^{-1/2}$, as shown in the seminal work of Saffman and Taylor[188]. The most unstable wavelength, the maximum amplification, can be described as $\lambda_m^* = \sqrt{3}\lambda_c^* = \pi W(\alpha^2 Ca(1-M))^{-1/2}$. The flat interface deforms into fingers which grow and compete to yield as single, symmetric finger that propagates steadily and occupies half the width of the channel for sufficiently large values of the product $\alpha^2 Ca$. The finger shape is captured by the steady Saffman-Taylor solution in the absence of surface tension[188], but the selection of the half-width finger requires the introduction of surface tension as a singular perturbation. Beyond a threshold value of $\alpha^2 Ca$, which depends on the level of roughness in the channel, experimental observations show that the advancing finger becomes unstable, undergoing repeated tip-splitting and forming disordered, highly branched patterns[189, 190]. The onset of disordered dynamics bears the hallmarks of a subcritical transition, and thus, finite-amplitude perturbations are presumably required to initiate complex time-dependent dynamics[191-193] (Fig. 6B). Subcriticality is a feature shared with the transition from laminar flow to turbulence in linearly stable wall-bounded shear flows, where weakly unstable states can orchestrate complex transient dynamics[194].

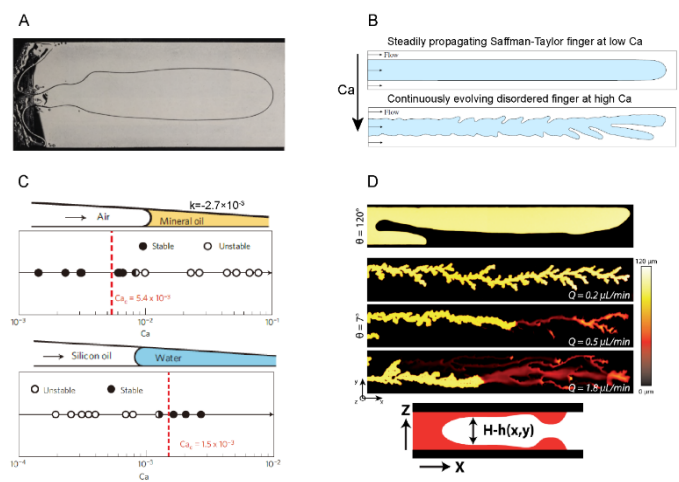

**Fig. 6.** Multiphase flow dynamics in Hele-Shaw (HS) experiments. (A) The seminal Saffman-Taylor viscous fingering experiment (128 cm long HS cell)[188] (Adapted with permission. Copyright 1958, The Royal Society (U.K.)). (B) Schematic diagram of subcritical transition from single finger to disorder[193]. (C) Inhibiting or triggering interfacial instability in the converging centimetric Hele-Shaw model[195] (Adapted with permission. Copyright 2012, Springer Nature. (D) Interfacial instability transition from drainage to imbibition (microfluidic experiments)[196] (Adapted with permission. Copyright 2014, American Physical Society).

The Saffman-Taylor finger can be further influenced by structural heterogeneity and wettability. For instance, even a simple gradient in the flow passage of a Hele-Shaw cell can lead to fundamentally different displacement behaviors[195], as illustrated in Fig. 6C. In a converging Hele-Shaw cell, when a low-viscosity fluid displaces a high-viscosity fluid, linear instability of the interface and fingering are suppressed to larger values of $Ca$. Conversely, when a high-viscosity fluid displaces a low-viscosity fluid, fingering can be triggered, with the flow regime transitioning from unstable to stable as $Ca$ increases[195]. This stability can be described using the following criterion:

$$1 - M - \frac{2k\cos\theta}{Ca} \leq 0 \qquad (13)$$

Here, $k = dH/dx$ represents the slope of the linear variation in cell depth along the direction of fluid motion $x$, and $\vartheta$ is the equilibrium contact angle of the invading fluid. The critical capillary number, $Ca_c = 2k\cos\theta/(1-M)$, denotes the point at which the interface undergoes a stability transition. Wettability is unlikely to influence the main findings of the Saffman-Taylor experiment in the absence of a depth gradient ($k=0$), because wettability effects become significant only when a gradient is present, as expressed in Eq. 13. However, the classical Saffman-Taylor experiment warrants revisiting when transitioning from drainage-dominated conditions to strong imbibition scenarios. Unexpected patterns can emerge in the forced imbibition regime where thin liquid films are entrained from the moving meniscus[196]. This occurs under high-$Ca$ conditions where a low-viscosity fluid displaces a high-viscosity fluid, as shown in Fig. 6D. The Hele-Shaw prototype has also been used to explore other boundary conditions, such as elastic boundaries[15, 173, 197], gap expansion[198], roughness[199], which significantly enrich our understanding of the conditions under which the transition between stable and unstable viscous fluid displacement is triggered or inhibited.

**Straight Channels.** Compared with the large aspect-ratio limit of the Hele-Shaw model, the propagation of an invading fluid finger into a defending fluid-filled tube of polygonal cross-section is a fully 3D configuration. Viscous instability is strongly influenced by the channel geometry, including shape factors and the aspect ratio. Additionally, capillary displacement can further complicate the flow processes due to the presence of corners or wedges in the channel cross-section. At a low $Ca$, the pressure drops across the fingertip and relative finger widths decrease with increasing $\alpha$ and $Ca$, but at sufficiently high $Ca$, the width reaches a minimum and begins to increase[200], as shown in Fig. 7A. In contrast to the stable, steadily propagating finger observed in a perfect square channel during drainage, perturbations in channel depth can introduce multiplicity of steady states and even periodic modes of viscous fingering[201]. This variable-depth system becomes increasingly sensitive to depth perturbations as the aspect ratio $\alpha$ increases. The bifurcation diagram in Fig. 7B, which shows the finger width (and speed) as a function of Ca, indicates the presence of unstable double- or triple-tipped modes (dashed lines), while solid lines correspond to stable states.

By contrast with the main meniscus deformation in the drainage regime ($\theta > 90°$), controlled by the viscous ratio $M$ and capillary number $Ca$, imbibition processes ($\theta < 90°$), especially the strong imbibition regime ($\theta < 30°$), present intricate capillary phenomena due to the competition between the main meniscus displacement, corner flow, and film flow in 3D microchannels. When $M \gg 1$, viscous instability does not occur. Under the constant flow rate, the imbibition processes in microchannels can be treated as classified as: classical capillary rise (main meniscus flow) in a right circular cylinder with the wetting fluid of contact angle $\theta$; compound capillary rise (coupled main meniscus flow and corner flow) in a right square tube with interior corner flow; and pure interior corner flow[202], as shown in Fig. 7C. When the imbibition length becomes significantly extended, the impact of inertial forces can be disregarded, characterizing this phase as the viscous regime. For classical main meniscus flow in a right circular cylinder without external pressure drop, by assuming Hagen-Poiseuille flow in the tube and negligible contribution of gravitational forces, the displacement of the main meniscus can be accurately described using the well-established Washburn equation[203]:

$$l = \sqrt{\frac{\gamma \cos\theta\, R}{2\mu} t} \quad (14)$$

where $t$ is the imbibition time, $R$ is the radius of the channel, and $l$ is the length of the capillary rise (i.e., the displacement length of the fluid-fluid interface). This equation yields the familiar "diffusive" result $l \propto t^{1/2}$. However, for microchannels with axial variations, microfluidic experiments, and theory have demonstrated that shape variations of the channel in the flow direction modify this 'diffusive' response (Fig. 7D). At short times, the shape variations are not significant, and the imbibition is still diffusive $l \propto t^{1/2}$. However, at long times, different power-law responses occur, and their exponents are uniquely connected to the details of the geometry[204]. For example, for conical channels, it yields $l \propto t^{1/4}$.

Consider a rectangle channel, the invading fluid's area saturation $S$, (i.e. the fraction of the cross-section area occupied by the invading fluid), and the spreading of the corner flow can be analytically expressed by solving the following equations[202, 205]:

$$\frac{\partial S}{\partial t} = \frac{1}{2\sqrt{C}} \frac{\gamma}{\mu\beta} \frac{\partial}{\partial x}\left(S^{1/2} \frac{\partial S}{\partial x}\right) \quad (15)$$

where the shape factor $C = 4(\cos\theta \cos(\pi/4 + \theta)/\sin(\pi/4) - (\pi/4 - \theta))$, $\theta$ is the contact angle, $t$ is the imbibition time, $\gamma$ is the fluid-fluid interface tension, and $\mu$ is the invading fluid viscosity. $S_0$ is the area of the wetting fluid at the junction of the corner flow and main meniscus. $S_0 = R^2 C/Wd$, where $R$ is the hydraulic radius at the junction of the corner low and main meniscus ($S = S_0$), $R = Wd/(W+d)\left(\frac{\cos\theta - \sqrt{(\pi/4-\theta)+\sin\theta\cos\theta}}{\theta - \pi/4 + \cos^2\theta - \sin\theta\cos\theta}\right)$. The dimensionless flow resistance as a function of contact angle θ and cross-section shape of microchannels is $\beta = C/[2\hat{g}_w(\cos\theta - \sin\theta)^4]$, where $\hat{g}_w$ is the dimensionless conductance of the wetting phase proposed by Patzek & Kristensen[206, 207]. A class of self-similar solutions was introduced by Dong et al.[205] for a nonlinear diffusion equation (Eq. 15) of rectangular microchannels. Therefore, it can also be derived that the front of spreading corner flow distance $x_f$ follows the rule $x_f \propto t^{1/2}$.

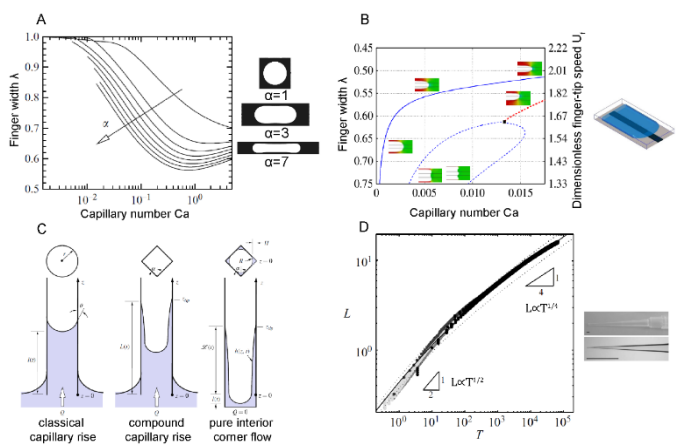

**Fig. 7.** Multiphase flow dynamics in straight microchannels. (A) Aspect ratio effect on the steady propagation of an air finger into a rectangular tube[200]. (B) Channel-depth perturbation effect on the viscous instability[201]. (C) Imbibition process in microchannels with interior corners[202] (Adapted with permission. Copyright 2012, Cambridge University Press). (D) Axial variation effect on main meniscus flow in microchannels[204] (Adapted with permission. Copyright 2008, Cambridge University Press).

**Pore-throat channels.** The pore-throat channel, i.e., a tube (either cylindrical or of other cross-sectional shape) with a constriction in the middle of its length, is a typical geometric unit in porous media. It has been widely utilized for fundamental studies of subsurface flow processes. In 1970, Roof introduced a pioneering quasi-static criterion for the snap-off of oil droplets in water-wet pore-throat channels with circular cross-sections[208]. According to this criterion, snap-off, which is the separation of the non-wetting droplets by connection between the wetting films present on the surfaces of the channel, occurs when the local capillary pressure at the throat exceeds that at the pore body. This study has since become a cornerstone for understanding snap-off behavior in both strong imbibition and drainage processes. It is worth noting that snap-off usually occurs in a strong imbibition or strong drainage to ensure that a corner flow or water film can be produced, the detailed critical wetting condition depends on the geometry of the channel[209]. For strong imbibition $\theta < (\pi - \varphi)/2$, while for strong drainage the contact angle must verify the condition $\theta > (\pi + \varphi)/2$, where $\varphi$ is the corner angle of channel cross-section. For example, because the cross-section of channels in microfluidic chips is generally rectangular ($\varphi = \pi/2$), snap-off will only occur if the contact angle satisfies $\theta < 45°$ or $\theta > 135°$.

For drainage processes, Fig. 8A illustrates the behavior in a constricted cylindrical capillary tube, where the non-wetting fluid advances through the constriction in a finger-like pattern, leaving behind a film of wetting fluid[210]. This fluid film thickens rapidly near the constriction at the tube's center, ultimately leading to snap-off events. The rapid growth of the fluid film is primarily influenced by the local capillary curvature and the displacement flow rate. Capillary experiments indicate that such phenomena in shrinking circular tubes can occur at capillary numbers as low as ~$10^{-5}$ during drainage processes[210]. However, snap-off events in rectangular channels are more complex due to the intricate geometric curvature controlled by the channel's width and depth, as depicted in Fig. 8B. For pore-throat junctions in 2D microchannels with rectangular cross-sections, the geometric criteria for capillary snap-off can be categorized into different types, depending on the relative ratio of channel depth to the curvature radius of the pore and throat[211].

For imbibition processes, snap-off events are similar to those in drainage processes, but they include an additional fluid film spreading process. The corner flow in a square channel is described by Eq. 15, and the critical capillary number $Ca_c$ for snap-off phenomena is derived from the velocity ratio of the corner flow to the main meniscus flow[16]:

$$Ca_c = \sqrt{2/3}\, a(K\beta\tau)^{-1/2} \qquad (16)$$

The first requirement for snap-off is Ca<$Ca_c$, meaning that when the corner flow velocity is significantly larger than the main meniscus flow velocity, sufficient corner flow can trigger snap-off during imbibition. This necessary condition has been validated through pore-scale simulations and microfluidic experiments (Fig. 8C). Here $\beta = C/[2\hat{g}_w(\cos\theta - \sin\theta)^4]$ represents the dimensionless flow resistance as a function of the contact angle. Other parameters are constant, such as $a$, which is a parameter introduced by the trial function, and the parameter $K$, which is calculated by a variational method[212]. For a tube with rectangular corners, $a$ and $K$ are 0.59 and 1.447, respectively. Here $\tau = 10^4$ is the dimensionless time for snap-off phenomena based on experimental observation[213, 214].
However, wettability has only a weak influence on the critical capillary number. For example, even if the contact angle varies from 0° to 40°, the critical capillary number decreases by less than an order of magnitude[141]. When the corner flow velocity exceeds the main meniscus flow velocity, and the capillary force at the throat is greater than that at the pore, the corner flow associated with the air-water configuration becomes unstable at the throat. Any small perturbation can force the system to merge the separate precursor interfaces, causing the water meniscus to snap across the cross-section and split the defending phase. The geometrical snap-off criterion for corner flow is given by[16]

$$\alpha_{p-t} > min(R_{asp}^t, 1)\frac{(R_{asp}^t + 1)}{(1 - tan\theta)R_{asp}^t} \qquad (17)$$

This criterion is controlled by the throat aspect ratio $R_{asp}^t = W_t/d_t$, the pore aspect ratio $R_{asp}^p = W_p/d_p$, and the depth variation factor $\alpha_{p-t} = d_p/d_t$, as illustrated in Fig. 11D. $W_p$ and $W_t$ are the width of the pore and throat, respectively. $d_p$ and $d_t$ are the width of the pore and throat, respectively. Theoretical analyses reveal interesting insights: microfluidic chips with a uniform depth significantly reduce the probability of snap-off. For example, chips fabricated using deep reactive ion etching (DRIE) on silicon or soft lithography on PDMS, where $\alpha_{p-t} = 1$, exhibit minimal snap-off events, as shown in the gray area of Fig. 8D. Similarly, snap-off phenomena are nearly impossible to observe in chips fabricated via isotropic etching methods, such as HF etching on glass, where $\alpha_{p-t} = 1$ and $R_{asp}^p$ & $R_{asp}^t > 2$. This explains why these unstable imbibition behaviors are rarely observed in traditional 2D microfluidic chips or 2D simulations with uniform depth.

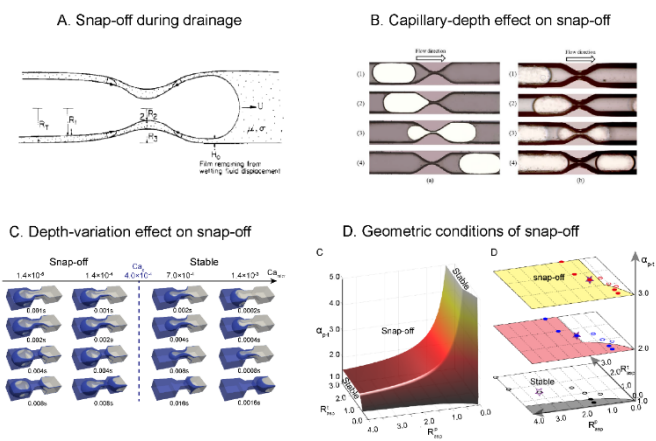

**Fig. 8.** Multiphase flow dynamics in the pore-throat channel. (A) Film deposition and snap-off in a constricted capillary during drainage[210] (Adapted with permission. Copyright 1988, American Chemical Society.). (B) Displacing results in constricted square capillary tubes with different depths: complete displacement for a depth of 16 microns (left) and snap-off for a depth of 25 microns (right)[211] (Adapted with permission, Copyright 2021, American Geophysical Union). (C) The critical capillary number for the snap-off phenomena during imbibition[16]. (D) The geometric criterion for snap-off phenomena in the $\alpha_{p-t} - R_{asp}^p - R_{asp}^t$ space based on theoretical analysis; (left) the color map represents the value of $\alpha_{p-t}$. (right) slices of the simulation and experimental data for stable and unstable snap-off events under different depth variation [16].

**Pore-doublet channels.** The parallel pore doublet model represents a simple network unit composed of two parallel capillaries and consists of three main components: a feeding channel that supplies the invading fluid; two capillary tubes that bifurcate upstream and rejoin downstream; and an exit channel. The pore-doublet model has been widely used as an idealized representation of pore structures to interpret the trapping of one phase by another immiscible phase during immiscible displacements in permeable porous media. For instance, bypass events arise from the velocity mismatch in the main meniscus displacement between two capillaries. The motion of fluid-fluid interfaces in pore doublets can be visualized using transparent capillary microfluidic chips. Assuming the wetting and non-wetting fluids have identical viscosities to simplify viscosity effects on pressure drop, Chatzis and Dullien[215] derived explicit velocity formulations for each capillary and provided a semi-quantitative understanding of fluid flow through a relatively long series of pore doublets whose two tubes have different widths. It can be concluded that the entrapment of the wetting phase is solely determined by the pore structure (Fig. 9A), whereas the entrapment of the non-wetting phase is influenced by the pore structure, the relative magnitude of viscous forces to capillary forces, and strong imbibition-induced film flow (Fig. 9B). To explore the influence of the viscosity ratio $M$ on the imbibition process (Fig. 9C), a theoretical analysis for a wide range of $M$ values spanning $10^{-4}$ to $10^3$ illustrates which of the two menisci reaches breakthrough first, depending on Ca and M. Typical regions are defined by the competition between viscous and capillary forces: Below the solid blue line in Fig. 9C, the meniscus in the narrow channel outpaces that in the wider channel at breakthrough, while above the solid blue line, the meniscus in the wider channel outpaces that in the narrower channel to reach breakthrough[216]. In porous media, what happens in geometric configurations similar to the pore-networks model with separate main menisci is critical, but the configuration of the parallel channels where narrower channels surround wider ones are equally important; they are, denoted as *merged pore-doublet* channels. These configurations focus on capillary competition in the pore-doublet but ignore the viscous effects, which originate from the situation in porous media where fluid-fluid interfaces merged frequently in parallel pores (Fig. 9D). Here the narrower channel, acting as an interior edge of wider channel, influences capillary interfacial stability. If the capillary pressure difference between the narrower and wider channels' meniscus is negative $\Delta P_c < 0$, the interface in the smaller channel invades preferentially, causing a bypass phenomenon[141]. Key parameters include the critical contact angle $\theta_c$ for bypass phenomena and the dimensionless geometry factor $\Psi$ of the parallel channels. Larger depth variations α lead to higher values of $\Psi$, which significantly affect the flow behavior.

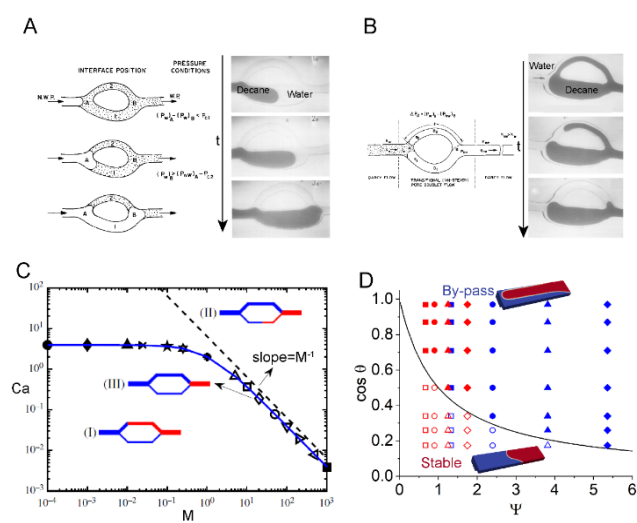

**Fig. 9.** Multiphase flow dynamics in pore-doublet channels. Conventional pore doublet model representation of (A) drainage-type displacement and (B) imbibition-type displacement[215] (Reproduced with permission, Copyright 1983, Elsevier). (C) *Ca-M* diagram showing the imbibition preference in a two-dimensional pore doublet (Reproduced with permission, Copyright 2021, Cambridge University Press). (D) Increasing the contact angle $\theta$ over the critical value $\theta_c$ will inhibit the bypass event[141] for the merged pore-doublet channel.

**Dead-end pores.** In contrast to the previously discussed structured microchannels, where larger viscous pressure drops can overcome capillary trapping[135, 136], mobilizing trapped fluid from dead-end pores presents a significant challenge. These dead-end pores are small, confined regions with only one exit, making it extremely difficult to force a second fluid through them. It was demonstrated that nanoparticles can modify the morphology of the solid surface, which can reverse the capillary trapping and lead to the self-removal of non-aqueous fluid from dead-end structures. To investigate multiphase flow dynamics in such environments, a microfluidic approach was developed, incorporating dead-end structures connected to a main flow channel to induce non-aqueous fluid trapping. Fig. 10A illustrates a typical dead-end structure with a microscopically rough surface and nanoscopic textures formed by nanoparticle suspensions. The experiments were conducted under weakly water-wet conditions, where spontaneous film formation could be ignored. Results demonstrate that nanoparticle suspensions can successfully trigger the release of non-aqueous fluid from dead-end pores (Fig. 10B). Compared with smooth microscopic convex surfaces where only a stable molecular adsorption film persists due to hydrodynamic film rupture, nanoscopic concave surfaces can induce the formation of a hydrodynamic film by capillary condensation (Fig. 10C). Once a hydrodynamic film is established, the capillary pressure gradient within the dead-end pore structure drives the displacement of non-aqueous fluid out of the pore (Fig. 10D).

Chemical reactions and solute transport processes can also significantly impact the flow behavior in dead-end pores due to strong concentration gradients between the pore interior and exterior. These processes include liquid-liquid extraction[217], fluid-fluid dissolution or emulsification[218, 219], salt precipitation[18], and diffusiophoresis of colloids[220, 221], among others. Progress with these physicochemical processes will be discussed in Sections 4.3 and 4.4.

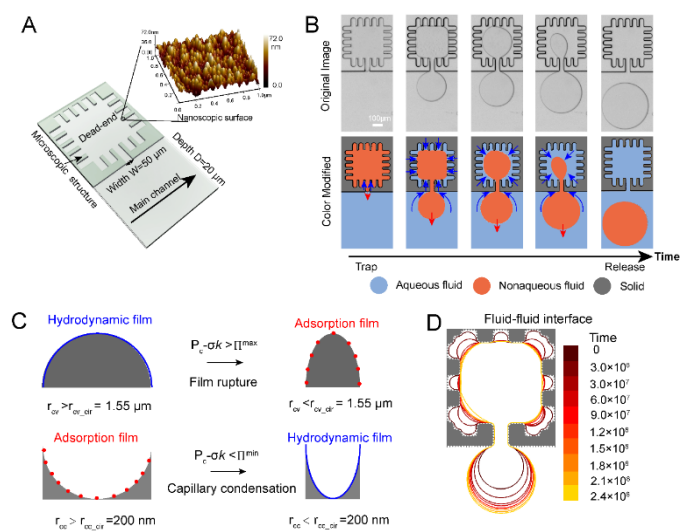

**Fig. 10.** Multiphase flows in dead-end channels. (A) Schematic illustrating the dead-end structure of the microchip, along with AFM images showing the nanoscopic structures formed by nanoparticle adsorption. (B) Experimental observation of the spontaneous release of trapped non-aqueous fluid from the dead-end structure. (C) Schematic diagram depicting the rupture of the hydrodynamic film to an adsorption film due to the small curvature radius of the microscopic convex surface, and the transition from an adsorption film to a hydrodynamic film driven by capillary condensation at the nanoscopic concave surface. (D) Evolution of the fluid-fluid interface in the dead-end channel driven by the capillary pressure difference. The release process from the dead-end, which depends on the stability of the hydrodynamic film, was analyzed using numerical simulations of flow based on the pseudo-potential lattice Boltzmann method.

**II. Multiphase flow patterns in microfluidic porous media**

As presented above, structural microfluidic channels capture fundamental interfacial phenomena that allow understanding local flow as well as interface displacement and instability effects in porous media flow, but they fail to describe multiphase flow patterns formation in porous media because they cannot account for the collective behavior of many meniscus interacting through the pressure fields in the two fluids, nor, more generally, for the properties of engineered and natural porous media, such as network topology, pore-size distribution, structural or wettability heterogeneity. Hence, microfluidic porous media, including millifluidic models, offers a more effective platform for studying porous media flow in such complex geometries consisting of a large number of interacting pores.

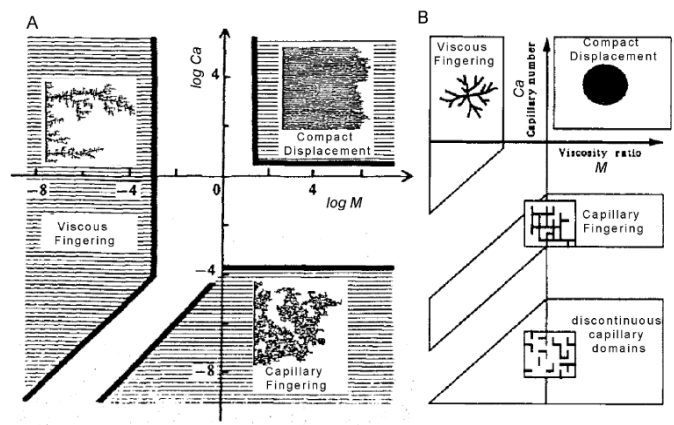

**Fig. 11.** Lenormand phase diagram for two-phase flow in 2D disordered porous media, showing the flow regimes as a function of the viscosity ratio M and capillary number Ca, in (A) drainage[184] (Reproduced with permission, Copyright 1988, Cambridge University Press) and (B) imbibition with narrow channels linking large pores[222] (Reproduced with permission, Copyright 1990, IOP Publishing).

When gravity can be neglected, for example, in the case of experiments in quasi-two-dimensional micromodels positioned horizontally, the two-phase displacement is governed by joint effect of capillary and viscous forces, whose relative influence is quantified by the capillary number *Ca*, viscosity ratio *M*, and contact angle *θ*. The evolution of displacement patterns in the porous matrix is highly contact angle-dependent for wettability between the extreme cases of fully nonwetting (strong drainage) and fully wetting (strong imbibition) conditions, because capillary forces render wider pores/channels easier to invade in drainage, while for imbibition the narrower pores/channels are more easily invaded. In a series of pioneering experimental studies, Lenormand and coworkers focused on fluid displacement in 2D microfluidic porous media with square lattices of rectangular capillary ducts with disorder in the channels' width[183, 222-224]. Although Lenormand mentioned the importance of wettability, he only explored the two limiting cases of a fully wetting and a fully non-wetting invading fluid[222]. For drainage, the seminal contribution by Lenormand, et al.[184] established the phase diagram of invasion patterns, which range from capillary fingering and viscous fingering to compact displacement (Fig. 11A). Capillary[225, 226] and viscous fingering[182] are two limits case where the displacement is controlled respectively by capillary and viscous forces, respectively. They are respectively observed synthesizes the results shown in Fig. 11A with a cross-over from stable (compact displacement) to unstable patterns (viscous fingering for M<<1 and capillary fingering for Ca<<1). It is important to note that, although pioneering studies from the 20th century employed millifluidic micromodels with characteristic sizes on the millimeter scale, the use of the relevant characteristic dimensionless associated with the aforementioned principle of similitude ensured that the underlying flow mechanisms became foundational for understanding multiphase flow in porous media. These findings on drainage have been widely observed by numerous subsequent microfluidic experiments[199, 227-229].

It is also interesting to note that for viscously unstable interface displacement, capillary fingering occurs at small scales, while viscous fingering is observed as large scale, with a crossover scale $l_c$ between these two regimes that varies a 1/Ca [230, 231]; in the limit cases of the phase diagram, the Ca is so low or so high that $l_c$ reaches either the system size (low Ca) or the pore size (high Ca) so that only capillary or viscous fingering is observed at all scales.

For imbibition, Lenormand[222] identified an additional regime with discontinuous capillary domains, to our knowledge the first published indication of the role played by film flow in imbibition; they also showed that the flow patterns could depend on the width of the pores relative to that of the channels linking the pores. As the system transitions from drainage to imbibition, Cieplak and Robbins[232, 233] identified three modes of motion for the main meniscus - burst, touch, and overlap. In this model, the interface between the invading and displaced fluids is represented by a collection of circular arcs whose radii are determined by the capillary pressure, which defines the pressure difference between the invading and defending fluids. A burst mode occurs when the curvature of a growing meniscus begins to decrease, causing a local drop in capillary pressure that drives the invading fluid to burst into the adjacent pore. A touch mode is triggered when the meniscus encounters an obstacle and assumes its equilibrium contact angle, which accelerates before reaching its maximum curvature. The overlap mode is a coalescence of neighboring menisci, which meet either at the three-phase contact lines or at the fluid-fluid interfaces. In contrast to the touch and burst instabilities, the cooperative nature of the overlaps effectively suppresses excursions of the invading fluid interface and thus leads to a smooth front. Holtzmann and Segre[234] incorporated the Cieplak-Robbins model into network models and demonstrated that this wettability effect stabilizes fluid-fluid displacement systematically (Fig. 12A). Displacement experiments in flat beds of glass beads[134, 235] and optically transparent quasi-2D microfluidic cells[236] further demonstrate that the crossover between a stable frontal displacement and capillary fingering of the invading fluid was observed for an advancing contact angle of about 90°, which fits surprisingly well with the Cieplak-Robbins model predictions based on the assumption of a quasi-static interface advance.

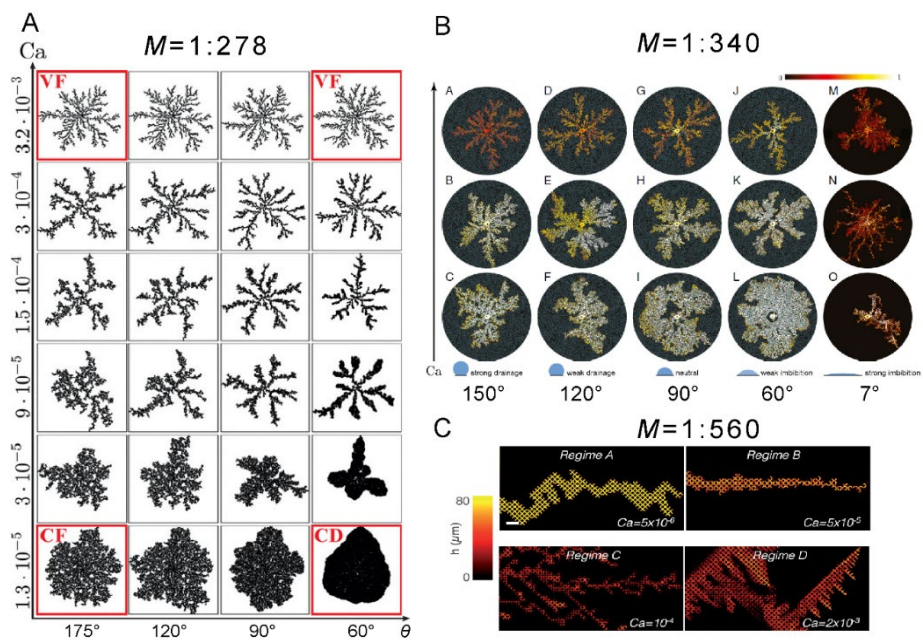

**Fig 12.** Wettability phase diagram from drainage to strong imbibition. (A) Invasion patterns simulated based on the Cieplak-Robbins model show that alternating wettability from strong drainage to weak imbibition can stabilize the displacement, which is controlled by capillary numbers $Ca$ and contact angles $\theta$ concerning the invading fluid. VF is viscous fingering, CF is capillary fingering, and CD is compact displacement[234] (Adapted with permission. Copyright 2015, American Physical Society). (B) Experimental displacement patterns show the invading fluid (water displacing silicone oil) at the breakthrough over the whole wettability range from strong drainage to strong imbibition[237] (Adapted with permission under CC BY-NC 3.0. Copyright 2016, National Academy of Sciences Publishing). (C) Four strong imbibition patterns illustrate the four regimes under different capillary numbers $Ca$ and extend the Cieplak-Robbins model to the strong imbibition regime. An aqueous solution is driven through a periodic lattice of channels filled with silicon oil. The color indicates the local thickness of the water films. (Adapted with permission. Copyright 2017, American Physical Society).

However, for strong imbibition, experiments have been found to deviate from the Cieplak-Robbins model for a large value of Ca and $M \ll 1$ due to a precursor film that spreads along the flat walls ahead of the main meniscus[237, 238]. This dependence of the emerging displacement pattern on the contact angle $\theta$ has been demonstrated comprehensively in microfluidic experiments[237] by varying the wettability of the solid from 0° to 180°, as illustrated in Fig. 12B. The corner flow under strong imbibition leads to incomplete displacement. For small invasion velocities, the characteristic formation of fingers of invading fluid is observed. In contrast to capillary fingers in the drainage regime, the fingers in this strong imbibition regime consist of chains of coalesced liquid rings around the base and the top of the posts. A similar phenomenon under strong imbibition has also been reported by Bartolo and colleagues in a flat microfluidic cell structured in a square network of rectangular ducts; the capillary number $Ca$ determines different imbibition scenarios that originate from two liquid-entrainment transitions and Rayleigh-Plateau instability[238], as shown in Fig. 12C.

In natural porous media with a wide distribution of pore and throat sizes, the 3D structure effect will be more obvious. For example, during both drainage and imbibition, the wetting fluid in the corner region of a throat can grow rapidly until the interface becomes mechanically unstable, resulting in spontaneous filling of the throat and disconnection of the nonwetting fluid[239]. However, microfluidic devices cannot easily capture 3D structural effects, which may ignore some important interfacial phenomena. To address the limitations of confined 2D models while retaining the ability to perform controlled visualization experiments, Lei et al.[16, 141] recently introduced an innovative fabrication technique, which leverages optimized sequential photolithography and multiple etching processes to create depth-variable microfluidic porous media. Experiments using these newly fabricated microfluidic chips revealed that in the regime of strong imbibition, the 3D pore geometry facilitates frequent snap-off events at low capillary numbers $Ca$ and high viscosity ratios $M \gg 1$, resulting in incomplete displacement of the defending fluid by the invading fluid. In contrast, snap-off events are suppressed in conventional 2D microfluidic porous media, indicating that interfacial phenomena within 3D pore geometries are more pronounced (Fig. 13).

The structure of porous media discussed above is random, thus capturing the statistical properties of natural porous media. However, microfluidic experiments have demonstrated that specific structural features—often relevant to sustainable energy applications—can significantly influence multiphase flow patterns. For instance, the synthesis of a periodic scaffold and its subsequent transformation into a self-organized liquid-infused material—where droplets spontaneously emerge with 3D microscale periodicity as a liquid within the porous scaffold—can significantly enhance carbon capture efficiency. This improvement is primarily due to the unprecedented precision and control over the interface area-to-volume ratio and perfusion[22]. A gradual and monotonic variation in pore size along the flow pathway[240] and the secondary pore structure in hierarchical porous media[241] can suppress viscous fingering during immiscible displacement. Additionally, preferential flow pathways in heterogeneous porous media can induce non-monotonic wettability effects by triggering corner flow and snap-off events in the matrix structure, which is far from the preferential flow pathway[132, 133]. Furthermore, incorporating increasingly complex interfacial characteristics, such as mixed wettability[242] and surface roughness[154], has demonstrated the diversity of mechanisms governing multiphase flow pattern formation.

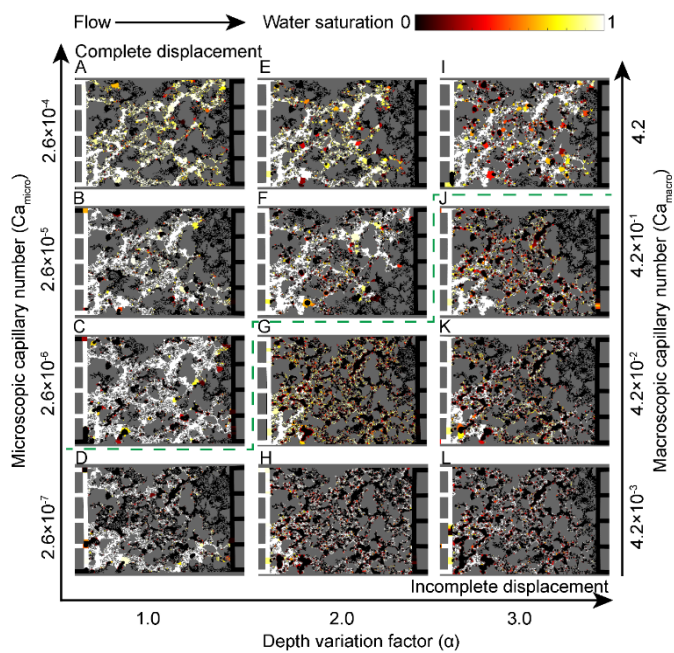

**Fig. 13.** 3D effect on imbibition phase diagram[16]. Imbibition patterns for microfluidic chips with different depth variation factors and capillary numbers. These patterns correspond to the breakthrough stage of the experiments when the displacing fluid reaches the outlet of the microfluidic chip. The color map shows the pore-average saturation of the displacing water.

When milli/microfluidic porous media are positioned horizontally, gravity does not play an important role, even for fluids with different densities. However, in subsurface applications such as $CO_2$ or $H_2$ storage, gravity may impact two-phase flow. In particular, any configuration where the denser fluid is positioned above the other one will make the fluid-fluid interface unstable, and conversely if the denser fluid is underneath[188]. Gravitational stabilization can for example cancel viscous finger growth and limit the roughness of the interface during capillary[243] or viscous fingering, with a roughness amplitude that scales as a decreasing power law of Bo-Ca[180].

### 4.3 Reactive transport in porous media

In recent years, the investigation of reactive multiphase flows has gained a lot of interest given its ubiquitous nature in energy subsurface applications. In the context of $CO_2$ sequestration in sandstone or deep saline aquifers[244], a number of microfluidics experiments[94, 157, 164, 245] were dedicated to the study of the dissolution of calcite by a $CO_2$ solution, providing insights into subsurface processes, e.g. where $CO_2$ min dissolved form reacts with the calcite, but may also exist as bubbles or a supercritical fluid and how these influence mineral dissolution and reactive surface availability. In this section, we will discuss the microfluidic experiments to study chemical reaction kinetics and its interaction with multiphase flow in porous media. Note that below we shall focus on heterogeneous reactions, that is, reactions between a solute species and the solid phase. However, that heterogeneous pore-scale flow in porous media limits mixing between solute reactants and thus controls situ reaction rates for homogeneous reaction at large Damköhler number[246] ; microfluidic experiments have also played a key role in understanding such effects.

**I. Interaction between reaction dynamics and single-phase flow in porous media**

Fluid-solid reactions in geological formations may facilitate the migration of environmentally significant fluids, such as $CO_2$ leaking through caprocks during geological carbon sequestration. When the flowing fluid is reactive, such as $CO_2$-acidified brine, fracture dissolution expands the aperture, forming diverse dissolution patterns that enhance fluid pathways. CT imaging can be used to analyze the effects of $CO_2$-acidified brine on fracture geometry[169, 247]. The permeability evolution of rock fracture is primarily driven by reaction-induced channelization rather than the overall extent of dissolution, which resulted in significant permeability enhancement[247-249]. To understand the reaction kinetics in fractures, microfluidic chips with optical microscopy can detect the transitions of dissolution patterns in the *Pe-Da* phase diagram[250, 251]. For example, injecting a NaCl solution flow through a NaCl crystal plate under varying flow rate and reaction rate conditions can mimic the simple reaction kinetics in porous media[251]. The critical Péclet numbers for transitions from compact to wormhole and from wormhole to uniform patterns increase with the reaction rate. To consider the impact of heterogeneous mineral composition, natural rock samples were embedded into microfluidic cells to probe multiscale dissolution dynamics. By combining high-speed imaging, scanning electron microscopy, and energy-dispersive spectroscopy, time-resolved images of the rock unveil the spatiotemporal dynamics of dissolution and illustrate the changes in the fracture interface[19]. Rock samples with strong mineral heterogeneity shows preferential reaction patterns in which large calcite regions are selectively dissolved, which yield the scaling law $\Gamma \propto \Psi^{1/2}$, where $\Gamma$ is the normalized solid–fracture interface length and $\Psi$ is the normalized fracture volume. However, this scaling law becomes $\Gamma \propto -\Psi^{4/3}$ for rock samples with homogenous minerals, such as carbonate-rich samples with large carbonate particles.

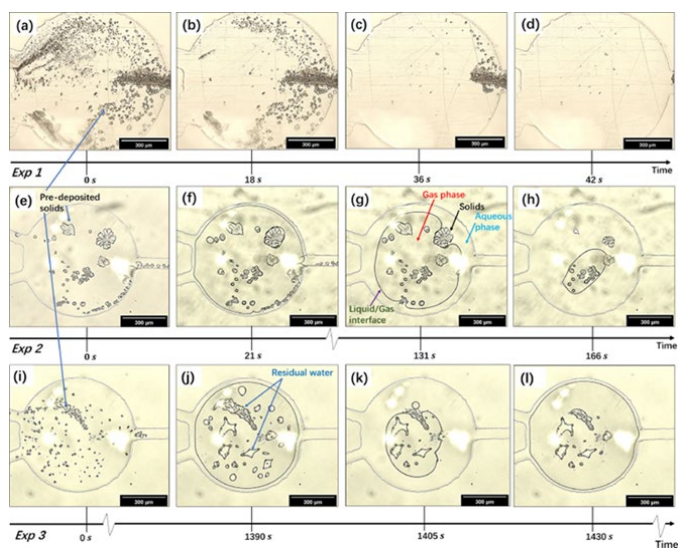

**Fig. 14.** Experimental snapshots of the three-dissolution regime during the course of the experiment of Xu and Balhoff[252] (Adapted with permission. Copyright 2022, Elsevier). The reactor chamber is circular and filled with crystals of calcite (marked as solid). An acidic solution injected from the two channels on the left and flows to the outlet through the channel on the right side.

To evaluate the effect of acid concentrations and injection rates on the dissolution process, Xu and Balhoff[252] conducted a controlled microfluidic experiment to induce the dissolution of calcite crystal in a circular chamber. The authors identified three distinct dissolution regimes namely:

Nongaseous Regime (observed at high Pe): This regime is characterized by high acid injection velocities where advection dominates the process. The injected rapidly dissolves the mineral surface without forming gaseous $CO_2$, resulting in linear and uniform dissolution patterns (see Fig. 14 a-d, Exp 1);

Gaseous Nonbreathing Regime (observed at intermediate Pe): At moderate injection rates, gaseous $CO_2$ bubbles form and stabilize due to capillary pressure. This stabilization slows dissolution by shielding portions of the mineral surface and leading to the gradual exposure of fresh surfaces as the reaction progresses (see Fig. 14 e-h, Exp 2);

Gaseous Breathing Regime (Pe = 4): At low injection rates, dissolution is influenced by cyclic expansion and contraction of $CO_2$ bubbles, driven by alternating phases of $CO_2$ supersaturation and dissolution. This regime follows a three-stage pattern: initial mineral dissolution, stabilization of the gas phase due to $CO_2$ buildup, and a transition back to liquid dominance as fresh surfaces are exposed (see Fig. 14 i-l, Exp 3).

These regimes were mapped on Pe and $Da_{II}$ diagrams, capturing the interplay between acid-mineral reactions and the influence of $CO_2$ gas bubble growth, decay, or movement in regulating the reactive surface availability of calcite for further dissolution. The identified regimes highlight the coupled effects of acid-mineral reactions, transport dynamics, and $CO_2$ phase behavior, providing a framework for predicting dissolution patterns under various injection conditions.

**II. Interaction between reaction dynamics and multiphase flow in porous media**

Dissolution processes may be strongly influenced by multiphase-flow effects, for example, if a reaction-induced gas phase forms a barrier that hinders contact between the acid and rock[253]. The role of the $CO_2$ gas transport was explored by Jimenez-Martinez et al.[156] The authors compared how single-phase and multiphase flow systems influence $CO_2$ gas-calcite reactions in a reacting pore network etched in a limestone rock (Fig. 15). The behavior of single-phase and multiphase flow systems differed significantly, particularly in terms of dissolution and flow dynamics. In the single-phase flow, dissolution was concentrated in high-velocity channels, often leading to localized "wormhole" formation due to a feedback loop between flow and reaction rates. These channels exhibited initially high dissolution rates (due to initial high surface roughness) but stabilized over time. Porosity and permeability increased in these areas of high flow velocity. By contrast, multiphase flow with supercritical $CO_2$ bubbles led to more homogeneous dissolution across the porous medium, and no wormholes were observed. Instead, the oscillation of bubbles in the channels dynamically rearranged the flow, enhancing fluid mixing, and constantly changing the pore water. This bubble-induced flow disruption redirects fluid into previously inactive dead-end channels, fostering calcite precipitation in areas where minimal reactions would have been observed in the case of a single-phase system. Although the presence of bubbles limits the total volume of brine in contact with the rock, the enhanced mixing and constant flow reorganization result in more uniform dissolution and porosity/permeability changes across the medium. These findings highlight how multiphase flow systems, with their dynamic fluid behaviors, offer significant advantages over single-phase systems in terms of promoting homogeneous reactions and facilitating better mixing in reactive flow through porous media.

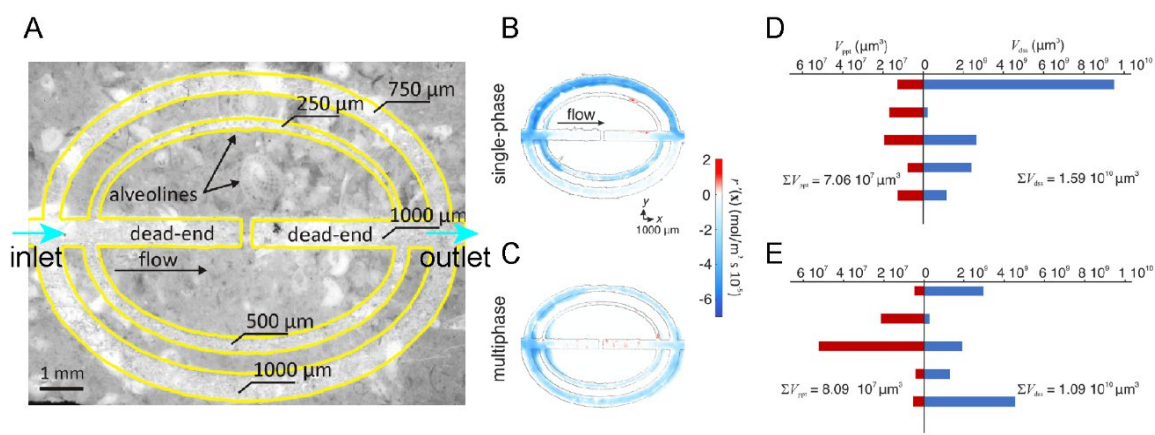

**Fig. 15.** Rock dissolution and mineral precipitation in single-phase and multiphase flow conditions[156] (Adapted with permission. Copyright 2020, American Geophysical Union). (A) Pore network design etched in limestone with the indicated inlet and outlet (B, C) Combined reaction rate (dissolution rate: negative values; precipitation rate: positive values) for both single-phase (B) and multiphase (C) flow experiments. (C, D) Dissolved ($V_{dss}$) and precipitated ($V_{ppt}$) volumes in each of the channels including dead-ends in the single-phase (D) and multiphase (E) experiments. The bars in panels (D) and (E) correspond to the vertical order channels (top to bottom) in panels (B) and (C), in the same vertical order. The total dissolved rock ($\Sigma V_{dss}$) and precipitated mineral volume ($\Sigma V_{ppt}$) are also provided for each experiment.

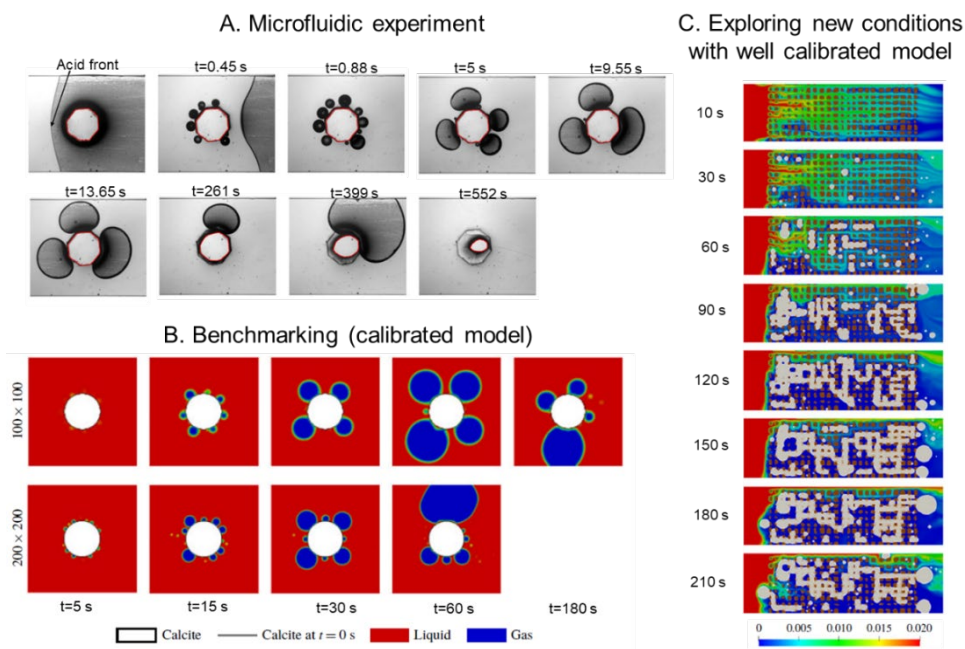

**Fig. 16.** The conceptual approach of benchmarking pore-scale models against microfluidic experiments[254] (Adapted with permission. Copyright 2018, Cambridge University Press). (A) micromodel showing calcite dissolution with $CO_2$ gas bubbles formation with time, (B) simulated results showing the effect of mesh discretization of the bubble on the reaction progress, and (C) simulations evaluating the inflow of acid solution in an array of calcite cylinders with development of $CO_2$ gas phase (grey) and the color bar indicated the acid concentration.

For reactive transport flows in porous media, the integration of experiments and models serves as a powerful approach for developing process-based models[255-257], which can subsequently be used to simulate more complex conditions. For example, Soulaine et al.[254] studied the dissolution of a calcite-carved cylinder in a microfluidic channel (Fig. 16), which served as the benchmark for the development of a Darcy-Brinkman model describing the dissolution processes with gas exsolution. The well-parameterized model was subsequently employed to simulate the fate of gas generated during mineral dissolution, focusing on the dissolution of an array of calcite cylinders representing a homogeneous reactive porous medium. Their numerical experiments revealed that while single-phase flow facilitated wormhole formation, the presence of gas impeded acid distribution, reduced dissolution rates, and inhibited the development of wormholes.

### 4.4 Biological effects in porous media

The pore-scale physics behind sustainable energy, such as underground hydrogen storage (UHS) and geological carbon sequestration (GCS) as the huge field-scale engineering application, porous materials for carbon capture, are much more complex, which involves interactions between microorganisms and multiphase flows. Recently, research has increasingly focused on understanding the biological effects on multiphase flow in porous media. For underground hydrogen storage, the biological effect is more obvious and significant that storage reservoirs like depleted oil and gas fields and brine formations are usually not sterile and contain a lot of microbial activities[55, 68]. $H_2$ serves as an electron donor in microbial processes such as methanogenesis, sulfate reduction, and acetic acid production. These processes, which often lead to adverse effects on storage and extraction efficiency, are influenced by the availability of microorganisms and electron acceptors (e.g., sulfate or carbon dioxide) within the reservoir. Elevated $H_2$ concentrations can stimulate the growth of hydrogenotrophic microorganisms. Various types of microorganisms in subsurface formations, including methanogens, sulfate-reducers, homoacetogenic bacteria, and iron(III)-reducers, thus play a significant role as hydrogen consumers[55]. Their activity is shaped by factors such as temperature, salinity, pH, and the availability of substrates. Dopffel et al.[68] summarized the microbial side-effects of underground hydrogen storage: gas mixture changes, souring, and $H_2S$ formation, steel corrosion by microbes and $H_2S$, microbial-induced clogging, dissolution of minerals and change in reservoir properties, and possible effects of hydrogen leakage. For geological carbon sequestration, microbiological pore clogging in reservoirs may be important for safety because bacterial plugs prevent $CO_2$ leaks[258]. Moreover, for carbon capture, liquid-infused materials could be used for living materials[22, 259], which can decrease $CO_2$ emissions, such as photosynthetic growth and lactate production by cyanobacteria[260]. The interplay between multiphase flow and biological activities will influence the transport of $CO_2$, nutrients, waste, and products in these biological multiphase systems. Therefore, monitoring and controlling microbial activity in the porous environment is essential. This section explores important advances in these sustainable energy solutions, organized by the underlying physical mechanisms.

**I. Interaction between biological activity and single-phase flow in porous media**

**Bacteria growth**. Bacteria adapt to diverse ecosystems and exhibit unique growth patterns influenced by environmental constraints. In laboratory settings, studies on two-dimensional (2D) planar surfaces have revealed a variety of morphologies, including circular colonies[261], herringbone patterns[262] and branched rough interfaces[263]. These patterns result from friction between the growing colony and the surface[264], as well as differential access to nutrients[265], influencing the overall function and physiology of bacterial communities, including resistance to antibiotics and parasites, resilience to environmental changes, and genetic diversity. By taking these factors into account, reaction–diffusion models[266], active continuum theories[262], and agent-based models[267] have been developed to explain these emergent colony morphologies. The real-world habitats are often three-dimensional (3D), such as gels and tissues within the host, soil and other subsurface media, wastewater treatment facilities, and natural bodies of water, which will influence the bacteria growth and lead to the different morphologies of colonies. Nutrients required by bacteria may also come from 3D geometric effects. However, due to the difficulty of experimental imaging, there have been relatively few studies on the three-dimensional morpho-dynamics of bacterial colonies. Recently, new colony morphologies were observed by using 3D granular hydrogel matrices[268]. They demonstrated bacterial colonies' transition to branched, broccoli-like morphologies under nutrient limitations. Unlike 2D growth, where nutrients are accessible from surrounding dimensions, 3D colonies experience internal nutrient depletion, driving surface instability and unique morpho-dynamics.

**Bio-clogging**. Pore clogging affects not only macroscopic parameters (i.e., permeability, heterogeneity), but also microscopic flow dynamics (i.e., preferential flow, pressure fluctuation)[13, 269]. According to the clogging mechanisms[270], it can be classified as: physical clogging by solid particles suspended in water, chemical clogging by chemical precipitation, and bio-clogging by the accumulation of microbial biomass in a porous medium. For bio-clogging, when high nutrient availability stimulates biomass growth, the direct accumulation of microbial biomass, including cells and extracellular polymers (EPS), can significantly decrease the hydraulic conductivity of the medium[271]. Bio-clogging presents significant technical challenges across multiple geoscience applications. These include groundwater extraction and various methods of artificial groundwater recharge, such as injecting water into the subsurface through wells, redirecting surface water via channels, utilizing infiltration basins or ponds, and employing irrigation techniques such as furrows or sprinkler systems. Two primary forms of bio-clogging were identified[272]: partial clogging of the pore is caused by the growth of immobilized biomass and detachment from or attachment to the pore wall, and complete plugging of the pore is caused by the growth of immobilized biomass and mobile cells filtration. It was demonstrated that biofilm streamers create rapid and severe flow disruptions in porous media without warning. Streamers significantly exacerbate clogging compared to wall-attached biofilms, with a timing determined by bacterial growth and clog duration influenced by cell transport and trapping[273]. By combining microfluidic experiments quantifying *Bacillus subtilis* biofilm formation and behavior in synthetic porous media with a mathematical model accounting for flow through the biofilm and biofilm poroelasticity, Kurz et al.[274] demonstrated that the closing of preferential flow pathways is driven by microbial growth, controlled by nutrient mass flow, while the opening of preferential flow pathways is driven by flow-induced shear stress, which increases as preferential flow pathways become narrower due to microbial growth, causing biofilm compression and rupture. Many theories have been proposed previously to describe the buildup of biofilm in porous structures[271, 275, 276].

**Hyperfluidic regimes due to collective motion of dense populations of pusher-type bacteria**. Populations of pusher-type bacteria (such as *Escherichia coli*) can reduce the viscosity of a suspension through a collective organization of their swimming, even inducing superfluidic regimes. It has been observed that in dilute conditions, for particularly active bacteria, the suspension exhibits a "superfluid-like" transition, in which the viscous resistance to shear disappears[277]. Further investigations into how this phenomenon depends on the size of the system, using bulk rheometry and particle-tracking rheological imaging, revealed the critical bacterial volume fraction of 0.75% required for the superfluidic phenomenon to occur[278]. Additionally, it was found that across all tested solvent viscosities (1–17 mPa·s), *Escherichia coli* can reduce the effective suspension viscosity to nearly zero[279].

**Fluid flow induces bacteria migration and dispersion, as well as heterogeneity in bacterial activity in behavior.** The

discussion above focused on the impact of bacterial activity on single-phase flow and fluid properties in the subsurface application. Conversely, fluid flow can exert a significant influence on bacterial activity. Liquids in underground porous structures are rarely static, as fluid flow is a fundamental characteristic of bacterial habitats in soils, aquifers, rivers, lakes, and even within animal and plant systems. Such flows transport nutrients, signaling molecules, and toxic compounds, generating velocity and nutrient concentration gradients that shape bacterial growth and behavior. Understanding how fluid flow affects bacterial activity is crucial for controlling bacterial activity and mitigating its effects on storage and extraction efficiency. Courad et al.[280] have summarized the influence of flow near surfaces, within channels, and through pores on bacterial dispersion, surface attachment, and biofilm formation.

Advances in microfluidic designs have expanded the scope of such studies by enabling quasi-2D complex porous geometries, which offer spatial contrasts in flow velocities and heterogeneous transport phenomenology at the pore scale, allowing researchers to simulate more realistic environments[281]. For instance, porous devices, characterized by surfaces with numerous tiny holes, provide conditions conducive to bacterial adaptation, growth, or migration. These designs are particularly valuable for investigating phenomena such as chemotaxis and rheotaxis.

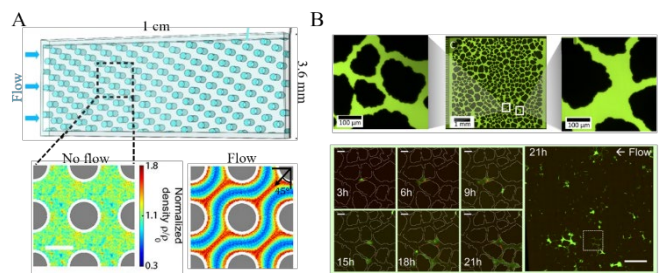

**Fig. 17.** Microfluidic devices featuring porous media geometries. (A) Comparison of the distribution of bacterial trajectories without flow and with flow condition in a homogenous microfluidic porous media chip[282], colored lines indicate density patterns, black arrows indicate mean flow directions (Reproduced with permission from the author). (B) Fluorescent images of Pantoea cells growing in the heterogeneous porous network. To mimic realistic surroundings, the network not only has constriction points smaller than a bacterium's cell size but also larger, highly connected pore spaces[283]. (Reproduced with permission under a CC-BY Creative Commons license, Copyright 2019, Public Library of Science).

To make analytical modeling easier, many researchers often use a homogeneous porous medium to simplify the model. For example, Dehkharghani et al.[282] employed a uniform microfluidic porous medium (Fig. 17A) to investigate the physical mechanisms governing bacterial dispersal in idealized porous flows. Comparing the results with those obtained under static conditions (i.e., without fluid flow), they observed that hydrodynamic gradients significantly restricted lateral bacterial dispersion while amplifying streamwise dispersion beyond the Taylor–Aris dispersion of passive Brownian particles. Their analysis further revealed that hydrodynamic reorientation of cells, combined with Lagrangian flow structures, generates filamentous density patterns.

Realistic environments are inherently heterogeneous, with heterogenous pore sizes. Therefore, researchers also try to design such heterogeneous porous structures. As shown in Fig. 17B, to mimic realistic surroundings, Aufrecht et al.[283] designed a porous network with not only constriction points smaller than a bacterium's cell size but also larger, highly connected pore spaces. They studied a wild-type Pantoea sp. isolated from the rhizosphere and its extracellular polymeric substance (EPS) knockout mutant. Both strains exhibited similar growth areas, with bacteria favoring specific routes until blockage occurred due to bacterial accumulation, causing fluid flow to redirect and supply nutrients to slower-growing bacteria. The key difference was that the EPS mutant formed long chains under flow conditions, whereas in static zones, the bacteria remained independent. Durham et al. also reproduced soil environments using a heterogeneous geometry and observed a tendency for *E. coli* to grow faster in close proximity to channels with higher flows[284].

Chemotactic responses of bacteria have also been shown to differ in porous and plain environments. For instance, *E. coli* cultured in a porous device with an α-methylaspartate (nutrient) gradient exhibited greater attraction compared to plain surfaces, suggesting that porous media flow enhances transverse migration in chemotactic bacteria[285]. Similarly, Listeria monocytogenes were studied in porous devices under acetate gradients, where acetate concentrations of 10–100 mM altered flagellar behavior, affecting motility and causing the bacteria to spin[286].

Further research highlights the role of fluid flow in driving phenotypic heterogeneity in bacterial growth and surface adhesion. Moreover, bacteria utilize mechanosensing to detect flow[287-290] and adapt surface adhesion by modifying bond types based on shear conditions[291-294]. Recently, Hubert et al.[265] found that the shear stress associated with velocity gradients near surfaces causes an *E. coli* population in the first stage of surface colonization (i.e. prior to biofilm formation) to induce phenotypic heterogeneity in the apparent division rates, with a proportion of the bacteria using their energy towards stronger surface attachment at the expense of cellular division; with increasing shear rate, the fraction of bacteria effectively dividing decreases. At a later stage of bacterial growth, during early biofilm formation, when the extracellular polymeric substances (EPS) matrix is underdeveloped, fluid flow still influences bacterial transport, attachment, and detachment, shaping spatial patterns and morphology[295, 296]. These processes strongly impact the eventual biofilm architecture. Conversely, in mature biofilms, EPS shields bacteria from flow-induced mechanical forces, with flow-biofilm interactions governed primarily by the mechanical properties of EPS. Moreover, bacteria utilize mechanosensing to detect flow and adapt surface adhesion by modifying bond types based on shear conditions. Recently, Wittig et al.[264] investigated the influence of shear stress on the shape, size, and distribution of microcolonies. By monitoring the three-dimensional spatial distribution of biofilms over seven days, they found that the biofilms consisted of smaller, pillar-shaped, microcolonies, with streamers emerging from the pillars' tips. While the shape, size, and distribution of these pillars depend on the imposed shear stress, this structure is seen as all shear stress values. Streamer formation is triggered by secondary flow induced by the base structure. Biofilm volume grew linearly over seven days across all shear conditions, with growth rates inversely related to wall shear stress. A scaling model provided insights into friction's role in limiting biofilm growth.

II. Impact of biological population on multiphase flow phenomena in porous media

Gas-liquid two-phase flow occurs naturally in the unsaturated zone (also denoted vadose zone) of the close subsurface, that is, between the Earth's surface and the free surface of unconfined aquifers. Two-phase flow processes also feature prominently in GCS, as the $CO_2$ injected at a depth below 900 m is in supercritical form, and displaces resident fluids upon injection into the geological formation[6]. Such residents can be brine, natural gas, or oil, depending on whether the formation is a deep saline aquifer or a depleted gas or oil reservoir. Two-phase flow introduces additional complexity into bacterial population behavior, due to interactions between bacteria and gas-liquid interfaces. These interfaces indeed affect microbial behavior, nutrient transport, and growth rates of the bacterial populations. Brizzolara et al.[297] investigated emulsification driven by immiscible Rayleigh-Taylor (RT) turbulence in marine oil spills. They observed that RT turbulence breaks the oil into smaller droplets within the water, increasing the interfacial area for a given oil volume. This enhanced interface promotes faster biodegradation of the oil by providing more surface area for marine bacteria colonization. Their findings emphasize the need to account for the entire emulsification process in deepwater spills or wave-driven overturning at the oil-water interface, rather than focusing solely on the initial instability phase. Using these different approaches to estimate biodegradation times could lead to significantly different environmental impact assessments. Liu et al.[298] studied how subsurface microbial growth impacts hydrogen storage through pore-scale experiments using halophilic sulfate-reducing bacteria. In a microfluidic pore network at 35 bar and 37°C, they observed significant $H_2$ loss from microbial consumption two days after injection. Over time, microbial activity and the consumption rate declined, which is strongly influenced by the surface area of $H_2$ bubbles and microbial activity. Microbial growth also altered the silica pore network's wettability from water-wet to neutral-wet, increasing the number of disconnected $H_2$ bubbles. These findings highlight critical factors affecting hydrogen recovery and injection rates and provide insights for assessing microbial risks and selecting suitable storage sites.

## 5. Linking pore-scale events to macroscopic flow behavior

Pore-scaling modeling, including multiphase flow, reactive transport, and microbial activities, is a critical tool for evaluating subsurface evolution in various sustainable energy-related applications. However, how to bridge these microscopic mechanisms with large-scale applications is a key challenge, i.e., effectively deriving constitutive equations that incorporate pore-scale heterogeneities into larger-scale analyses.

Microfluidic experiments can identify pore-scale multiphase flow and reactive transport phenomena at both single-pore scale and global porous media scale, which will support the development and validation of an upscaling method. In this section, we will introduce detailed how to link pore-scale events to macroscopic flow behavior in porous media.

### 5.1 Pore-network model for multiphase flow in porous media

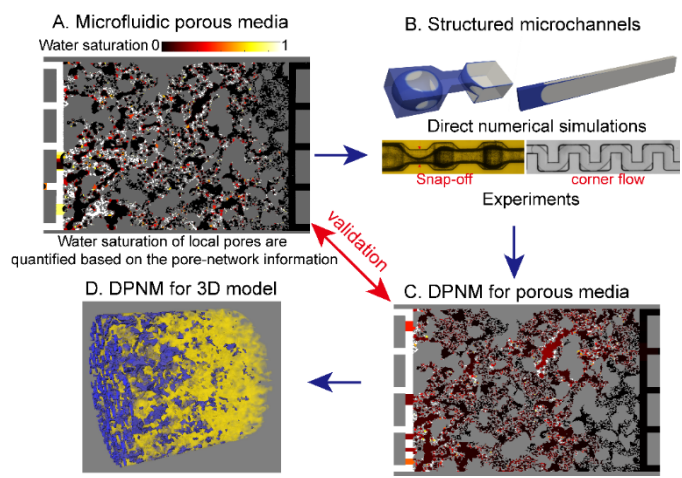

**Fig. 18.** Multiscale methodology by combining microfluidic porous media and numerical upscaling methods. (A) Visualizing and quantifying multiphase reactive flow processes in microfluidic porous media. (B) Characterizing interfacial or reactive transport phenomena at the single-pore scale by combining microfluidic experiments and direct numerical simulations in structured microchannels and building corresponding constitutive models. (C) Developing and validating the numerical upscaling method (DPNM) by incorporating constitutive models at the single-pore scale for transport phenomena in microfluidic porous media. (D) Predicting multiphase reactive flow processes in 3D engineered porous media.

It is nearly impossible to compute the multiphase flow in 3D-engineered porous media over a significant number of pores by direct numerical simulations due to huge computational costs. By incorporating important newly observed pore-scale events into the pore-network model, the multi-scale platform for upscaling multiphase flow phenomena from the pore scale to the global porous system scale for 3D-engineered porous media can be developed (Fig. 18). Firstly, macroscopic invasion patterns can be visualized and quantified by conducting microfluidic experiments with complex porous geometry (Fig. 18A). Then using a combination of microfluidic experiments in typical structured microchannels that model individual pores, direct numerical simulations, and theoretical analysis of the corresponding flows, it can be demonstrated that how pore geometry or surface properties affect interfacial dynamics (Fig. 18B). Based on these constitutive models of pore-scale events, a numerical upscaling method (dynamic pore-network model, DPNM) can be developed and validated (Fig. 18C). Here, pore-network models (PNM) are an intuitive and computationally inexpensive alternative for simulating porous media flows, where the porous geometry is simplified into interconnected pores and throats. Compared to quasi-steady PNM where interfacial events are determined by capillary forces alone, the dynamic pore-network model (DPNM) includes interfacial events controlled by both viscous and capillary forces[299]. Once interfacial events are updated based on local pressure and velocity through their constitutive models, the overall pressure and velocity field can be determined using the conservation equations (Kirchhoff's equations) [300]. The improved DPNM that incorporates novel constitutive models can be extended to 3D-engineered porous media and validated qualitatively against pore-by-pore images of multiphase reactive flow in microfluidic porous media or typical 3D-engineered porous media (Fig. 18D).

**5.2  Upscaling relationships in evolving porous media**

Reactive transport modeling is also important for evaluating subsurface evolution in various energy-related applications, which further complex the pore-scale constitutive model and transport in porous media. A key challenge for upscaling these reactive transport models is to accurately describe coupled processes, i.e., the interplay between chemical processes and the resulting changes in transport parameters, such as porosity and diffusivity. Commonly used empirical relationships, such as Archie's law, in which the pore scale molecular diffusivities of solute to the rock's porosity, and Kozeny-Carman equation, have been shown to fall short in addressing this complexity[257,301]. Therefore, there is a need to develop process-based relationships that integrate microstructural information into continuum-scale reactive transport models through upscaling.

Pore-scale simulations capturing the evolutions of the porous media over a wide range of Péclet and Damköhler numbers in combination with machine learning are foreseen as an efficient methodology[256]. For example, numerical investigations were conducted to evaluate how mineral precipitation patterns influence the transport properties of an inert porous medium and showed that the aforementioned Archie's law (classically employed in reactive transport models) was insufficient, prompting the development of an extended law incorporating critical porosity and effective diffusivity. However, questions persisted regarding the physicochemical processes captured by the extended law and their suitability for scenarios involving coupled mineral dissolution and precipitation. This was further investigated by Poonoosamy and co-workers combined time-lapse high-resolution optical microscopy with confocal Raman spectroscopy to investigate changes in evolving porous media due to precipitation processes[302]. In the first step, the microfluidic device featured a 2D pore network connected to dual supply channels, enabling counter-diffusive mixing of solutes that triggered mineral precipitation within the pore network[87]. As the pore network clogged, in-operando 2D Raman imaging visualized water transport through the evolving microporosity of the precipitates. This revealed the dynamic nature of porosity clogging and the resulting changes in effective diffusivity. Numerical tracer experiments using pore-scale modeling on 2D images of the evolving pore network further enabled the determination of effective diffusivity and the derivation of porosity-diffusivity relationships in line with the proposed extended equation. Building on this, they developed a second lab-on-a-chip experiment with a novel micromodel design to monitor the 3D evolution of porous media undergoing coupled mineral dissolution and precipitation processes driven by diffusive reactive fluxes[87] (Fig. 19). They investigated the replacement of celestine by strontianite, where a net porosity increase was expected due to the smaller molar volume of strontianite. However, under their experimental conditions, the accessible porosity and, consequently, the diffusivity decreased, reinforcing that Archie's law is not applicable and that the extended equation would be a more suitable approach.

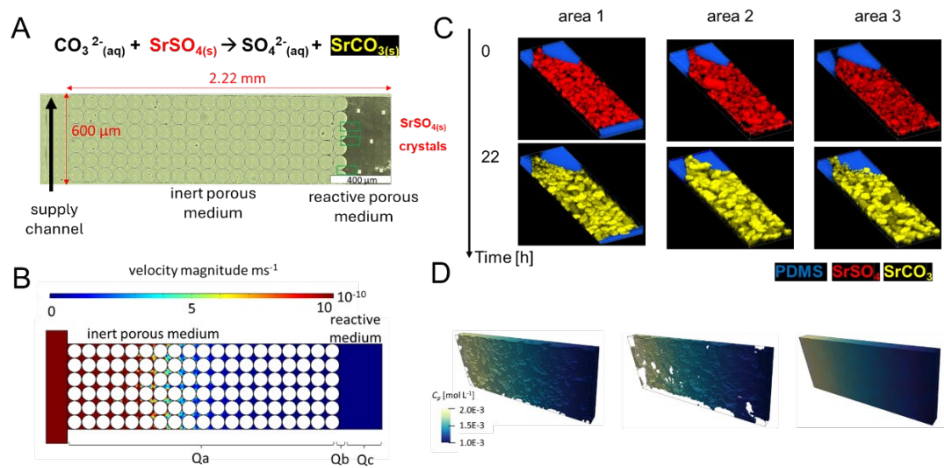

**Fig. 19.** Experimental and modelling results for investigating the effect of coupled mineral dissolution and precipitation on diffusive transport of solutes[87]. (A) micromodel concept with a supply channel of reacting solutes that diffuse in the inert porous medium and the reactive porous medium celestine (SrSO4) crystals. (B) Modelling of the velocity field in the microfluidic reactor using COMSOL Multiphysics indicating diffusive transport across the porous medium. (C) Initial and final Raman tomographs of the sampled areas and (D) 3D view of the simulated tracer concentrations across the sampled porous media in initial state of the unreacted porous medium (left), final state assuming no diffusion through the precipitates (middle), and final state with diffusion through the precipitates (right) with modelling results indicating no significant change in porosity.

These results highlight the importance of calibrating pore-scale models with quantitative microfluidic experimental data before performing simulations across a wide range of Péclet and Damköhler numbers. Such simulations can further inform the derivation of upscaled parameters. Future work in this field could integrate microfluidic experiments with automated image analysis and stream the results into a flow solver for deriving upscaled transport parameters, leveraging neural networks to enhance predictive capabilities (Fig. 20).

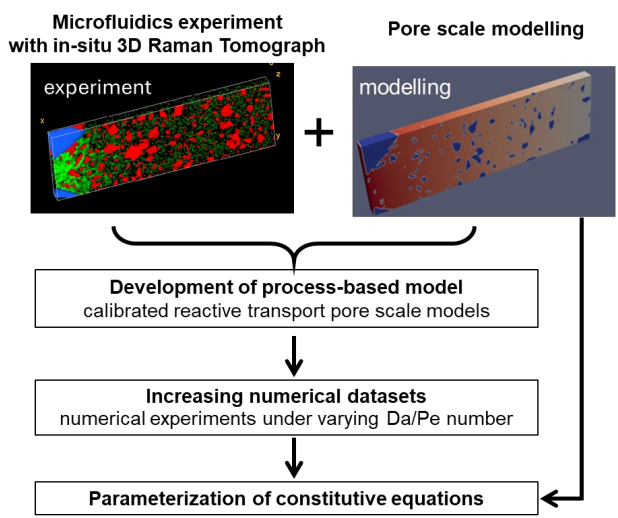

**Fig. 20.** A conceptual approach for developing rigorous constitutive equations quantifying the impact of mineral dissolution and precipitation on transport parameters.

## 6. Conclusion

Sustainable energy remains a central priority in global decision-making, with both industries and governments focusing on solutions to address the current energy and environmental challenges. While the time and length scales of microfluidic porous media are typically much smaller than those involved in industrial processes, their application offers significant potential in supporting a low-carbon energy future. This paper has presented a concise review of multiphase reactive flow mechanisms in porous media through microfluidic experiments, highlighting key scientific challenges in sustainable energy solutions. It has covered the fabrication and design of microfluidic chips, the complex interactions of multiphase flow, reactive transport, microbial activities, and strategies for upscaling pore-scale events to macroscopic flow behavior.

Microfluidics provides an excellent platform for observing, manipulating, and quantifying these intricate physico-chemo-biological processes in engineered porous structures. Such experiments enhance our understanding of pore-scale phenomena such as interfacial dynamics, reaction kinetics, and biological growth and migration within confined pore spaces. In doing so, they contribute to the multiscale integration of predictive tools for multiphase reactive flow in porous media. These insights are essential for describing the key transport, thermodynamic, and kinetic processes in sustainable energy systems. At a minimum, microfluidic porous media offers more accurate predictions of multiphase reactive flow behavior, laying the groundwork for their integration into sustainable energy solutions. Ideally, combining multiscale microfluidic experiments with upscaling methodologies could complement large-scale studies, optimizing the management of complex multiphase flows. With growing industrial investment in addressing these multiscale challenges, microfluidic porous media has the potential to become a pivotal platform for driving economic and environmental efficiency in the energy sector.

While significant progress has been made in understanding the fundamental mechanisms of multiphase reactive flow in porous media via microfluidic experiments, several challenges remain. Research on multiphase flow is relatively mature, while studies on reactive transport and microbial activities are still emerging. More basic research is needed to gain a comprehensive understanding of these complex processes. The integration of physico-chemo-biological processes presents substantial experimental challenges. Industrial porous media in energy systems are typically spatially heterogeneous and involve a wide range of physicochemical processes across multiple length scales. Despite advances in microfluidic porous media, these systems have yet to capture the full complexity of real-world processes. Ensuring the representativeness of diverse physico-chemical-biological interactions in microfluidic systems, and translating these findings to larger-scale applications, remains a significant challenge for future research.

## Conflicts of interest

There are no conflicts to declare.

## Acknowledgements

W.L. and S.B. acknowledge funding from the Wallenberg Initiative Materials Science for Sustainability (WISE) funded by the Knut and Alice Wallenberg Foundation and the European Union through the European Research Council under LUBFLOW (ERC-CoG-101088639) grant. S.Y. and Y.M. gratefully acknowledge funding by the French National Research Agency (ANR) under the "IMAGE" project (project number: ANR-21-CE04-0013). Y.Y. and J.P. acknowledge funding from the European Research Council through the project GENIES (ERC, grant agreement 101040341). A.J. acknowledges funding by the UKRI EPSRC research grant (EP/T008725/1). M. W. acknowledges the NSF grant of China (No. 12432013, 12272207) and the National Key Research and Development Program of China (No. 2019YFA0708704).